\documentstyle[psfig,amssymb]{mn2e}

\newif\ifAMStwofonts

\newcommand{\degrees}{$\rm ^{\circ}$}

\begin{document}

\title[2-D Kinematics to 3 R$_{\rm eff}$
]
{Probing the 2-D kinematic structure of early-type galaxies out to 3 effective radii
}
\author[
Proctor et al. 
]
{Robert N. Proctor$^{1,2}$, Duncan A. Forbes$^{1}$, Aaron J. Romanowsky$^{3}$, Jean P. Brodie$^{3}$, \\
\\
\LARGE Jay Strader$^{4*}$, Max Spolaor$^{1}$, J. Trevor Mendel$^{1}$, Lee Spitler$^{1}$\\
\\
$^1$ Centre for Astrophysics \& Supercomputing, Swinburne University,
Hawthorn VIC 3122, Australia\\
$^2$  Universidade de S\~{a}o Paulo, IAG, Rua do Matão, 1226, S\~{a}o Paulo, 05508-900, Brasil\\
$^3$ UCO/Lick Observatory, University of California, Santa Cruz, CA 95064, USA\\
$^4$ Harvard-Smithsonian Centre for Astrophysics, 60 Garden St., Cambridge, MA 02138, USA\\
$^*$ Hubble Fellow\\
Email: rproctor@astro.iag.usp.br, dforbes@astro.swin.edu.au}

\pagerange{\pageref{firstpage}--\pageref{lastpage}}
\def\LaTeX{L\kern-.36em\raise.3ex\hbox{a}\kern-.15em
    T\kern-.1667em\lower.7ex\hbox{E}\kern-.125emX}

\newtheorem{theorem}{Theorem}[section]

\label{firstpage}

\newpage

\maketitle

\begin{abstract}

We detail an innovative new technique for measuring the 2-D velocity
moments (rotation velocity, velocity dispersion and Gauss-Hermite
coefficients h$_3$ and h$_4$) of the stellar populations of galaxy halos using spectra from Keck DEIMOS
multi-object spectroscopic observations. The data are used to
reconstruct 2-D rotation velocity maps.

Here we present data for five nearby early-type galaxies to $\sim$3
effective radii.  We provide significant insights into the global
kinematic structure of these galaxies, and challenge the accepted
morphological classification in several cases. We show that between
1--3 effective radii the velocity dispersion declines very slowly, if
at all, in all five galaxies. For the two galaxies with velocity
dispersion profiles available from planetary nebulae data we find very
good agreement with our stellar profiles.  We find a variety of
rotation profiles beyond 1 effective radius, i.e rotation speed
remaining constant, decreasing \emph{and} increasing with radius.
These results are of particular importance to studies which attempt to
classify galaxies by their kinematic structure within one effective
radius, such as the recent definition of fast- and slow- rotator
classes by the SAURON project. Our data suggests that the rotator
class may change when larger galacto-centric radii are probed. This
has important implications for dynamical modeling of early-type
galaxies. The data from this study are available on-line.

\end{abstract}

\begin{keywords}
galaxies: general, galaxies: individual; (NGC~821; NGC~1400; NGC~1407; NGC~2768; NGC~4494), galaxies: kinematics and dynamics, galaxies; structure, stellar dynamics, techniques; spectroscopic, 
\end{keywords}

\section{Introduction}
Since the pioneering work of Simkin (1974), major advances in the
measurement of stellar orbits in early-type galaxies have been
made. The first quantities to be measured were the line-of-sight
rotation velocity (V$_{\rm rot}$; which can be considered a measure of
ordered motion within a galaxy) and velocity dispersion ($\sigma$; a
measure of the random motions). By combining these two measures with
photometric ellipticity ($\epsilon$) it was possible to probe the
anisotropy of stellar orbits in galaxies (Binney 1978).

Further progress was made in this field when, in the 1980s, Bender
(1988) showed that although the isophotal shapes of early-type
galaxies were well described by ellipses, small variations from
ellipticity could be used to categorise them as either `boxy' or `discy',
and that these variations relate directly to the stellar orbital
isotropy.  The combination of surface photometry and detailed
kinematics were therefore demonstrated to provide significant insights
into the global structure of galaxies.

More recently, higher order moments of the line-of-sight velocity
distribution (LOSVD) have been measured (e.g. Bender 1990; Rix \&
White 1992; Gerhard 1993; van der Marel \& Franx 1993). These 
characterise deviations from a pure Gaussian shape, e.g. skewed
and symmetric deviations, denoted h$_3$ and h$_4$ respectively.   
Such parameters can indicate kinematic substructures, such as embedded
stellar discs. 

The development of integral field units (IFUs), such as the SAURON
instrument, have given new impetus to this field, as these provide 2-D
maps of the LOSVD. This allows the definition and estimation of a
number of important new parameters that can be averaged over the
surface of a galaxy in a self-consistent way. Perhaps the most
important of these is the `spin' parameter ($\lambda_R$) which is a
proxy for the projected angular momentum per unit mass. Applied to the
SAURON galaxy sample of 48 early-type galaxies, it was used to show that a
dichotomy exists between `slow-' and `fast-' rotators (Emsellem et
al. 2007).  Trends between photometric and kinematic axis ratios and
position angles have also been shown to vary with $\lambda_R$; with
slow-rotators often exhibiting strong misalignments.  These
observations provide new clues to the formation mechanisms of these
differing galaxy types (Krajnovi{\'c} et al. 2008; Cappellari et
al. 2007; Thomas et al. 2007).

Statler \& Smecker-Hane (1999) used a different approach to building 2-D
maps of the LOSVD in NGC~3379. Using long-slit spectroscopy along four
position angles, they were able to reconstruct a model of the 2-D
kinematics by making suitable interpolations between their observed
axes. However, such studies are rare due to the large amount of
telescope time required to observe multiple position angles out to
large radii.

Despite these advances, most observations typically reach to less than
one effective radius, with only a handful of studies extending this
coverage up to 2--3 R$_{\rm eff}$ (but see Saglia et al. 1993; Graham
et al. 1998; Mehlert et al. 2000; Proctor et al. 2005;
Sanchez-Blazquez et al. 2006; Forestell \& Gebhardt 2008).  However,
such outer regions potentially provide key information regarding the
physical properties of galaxies. For example, radii within 1~R$_{\rm eff}$ 
only sample 50\% of the baryonic mass and $\sim$ 10\% of the
angular momentum of an idealised galaxy with an R$^{1/4}$ like light
profile and a flat rotation profile. The situation is even worse when
considering the distribution of total galaxy mass (including dark
matter). For instance, the simulations of Dekel et al. (2005) show
that only $\sim$5\% of the total mass is contained within 1~R$_{\rm eff}$ 
(see also Mamon \& Lokas 2005). Furthermore, the signatures of
some formation mechanisms only become apparent at large
galacto--centric radii (e.g.  McMillan, Athanassoula \& Dehnen 2007;
Weil \& Hernquist 1996).  A means to probe the kinematics in the outer
regions of early-type galaxies, and hence include a significant
fraction of a galaxies' angular momentum and mass, is therefore highly
desirable.

In this paper we measure stellar LOSVDs from deep multi-object
spectroscopic observations and develop a new technique for their
analysis. Specifically , we measure rotation velocity, velocity
dispersion and the Gauss-Hermite coefficients h$_3$ and h$_4$, out to
$\sim$3 R$_{\rm eff}$ in five nearby early-type galaxies.  By
extending the region studied to such radii we encompass more than 85\%
of the baryonic mass, 30\% of the angular momentum and 15\% of the
total mass, a significant increase in the fractions so surveyed in
galaxies. To achieve this we exploit the large aperture of the Keck
telescope, the high sensitivity of the DEIMOS instrument at 8500~$\AA$
and the strong calcium triplet absorption features at these
wavelengths. Even with this configuration, exposures of $\sim$~2~hours
are required for our analysis.\\

The paper is organised as follows: In Section \ref{odr} we outline the
observations and data reductions, detailing the technique by which we
extract the host galaxy spectra. In Section \ref{anal} we outline our
new kinematic analysis technique.  Results are presented in Section \ref{results}. We
discuss our findings in Section \ref{discuss}, and present our
conclusions in Section \ref{concs}.

\section{Observations and data reduction}
\label{odr}
\begin{figure}
\centerline{\psfig{figure=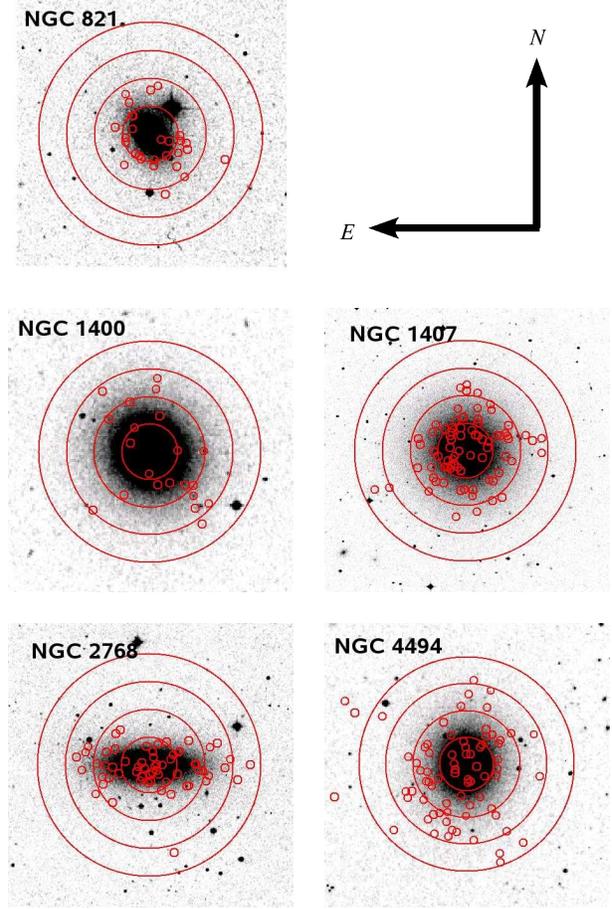,width=8cm}}
\caption{DSS images of the 5 target galaxies. Small red circles mark
the location of slits from which galaxy background spectra were
obtained. Large red circles represent 1, 2, 3 and 4 effective
radii.}
\label{thumbnails}
\end{figure}

\subsection{Sample selection}

The galaxies in our sample were selected with the intention of
obtaining spectra of their globular clusters (GCs). The galaxies are
all nearby, early-types with moderately rich GC systems. The selection
also had the requirement that recession velocities were such that the
calcium triplet lines would not generally be adversely affected by
strong sky lines. The main science goal was to obtain GC kinematics,
and hence probe the mass profile of their host galaxy halos. Initial
results for NGC 1407 from the 2006 Nov. run have recently been
reported in Romanowsky et al. (2008).

\subsection{Observations}
\label{data}
The data used in this work are from three observing runs using DEIMOS
on the Keck-2 telescope. The first observing run was carried out in
2006 Nov. using the 1200~l/mm grating centered at $\sim$7500~$\AA$ and a 1
arcsec slit. This allows coverage of the calcium triplet at 8498,
8542 and 8662~$\AA$ at a resolution of $\sim$1.5~$\AA$.  The second
run was carried out in 2007 Nov. and the third in 2008 April, both using
the same instrument setup as the first run. Seeing was $\sim$0.5 to
0.8 arcsec on all nights. A total of five galaxies were observed
(NGC~821, NGC~1400, NGC~1407, NGC~2768 and NGC~4494).

Between two and six independent masks were observed per galaxy and the
16 x 5 arcmin field-of-view allowed the observation of $\sim$80-120
spectra in each mask. However, only $\sim$20\% of these were at radii
from which the signal of the background host galaxy was sufficient for
the extraction of spectra (see Table \ref{fits}).

Fig. \ref{thumbnails} shows the positions of slits from which host
galaxy spectra were extracted.  We note that it was only possible to
measure recession velocities in some of the outermost slits,
i.e. velocity dispersion and higher order moments are not presented
for some of the outermost positions.

\begin{table*}
\begin{centering}
\begin{tabular}{|c|c|c|c|c|c|c|c|}  
\hline
Galaxy   & Hubble  & Distance&R$_{\rm eff}$   & M$_K$    &  PA$_{\rm phot}$   &Axis ratio & V$_{\rm sys}$ \\
         & Type  &  (Mpc)    & (arcsec)     & (mag)      &  (\degrees) &(K band)  & (km s$^{-1}$) \\
\hline	 		   		  	         	      		  
NGC~821  & E6?   & 22.4    &   50       & -23.6     &   30       & 0.62  & 1729    \\  
NGC~1400 & SA0-  & 25.7    &   29       & -23.8    &   35       & 0.90   &  574  \\
NGC~1407 & E0    & 25.7    &   70       & -24.9    &   60       & 0.95   & 1782   \\
NGC~2768 & S0$_{1/2}$& 20.8&   64       & -24.6    &   93       & 0.46   & 1327   \\
NGC~4494 & E1--2 & 15.8    &   49       & -24.8    &  173       & 0.87   & 1335   \\
\hline
\end{tabular}
\caption{Properties of galaxies featured in this work. Distances are
  from surface brightness fluctuations by Tonry et al. (2001), with
  the distance moduli modified by --0.16 according to Jensen et
  al. (2003). The values for NGC~1400 and NGC~1407 are taken as the
  average of their individual values.  Hubble types are from NED and effective
  radii are from RC3. Absolute K band magnitudes are calculated from
  2MASS apparent magnitudes and the distances given in column
  3. Photometric position angles (PA$_{\rm phot}$) and K band
  photometric axis ratios are from 2MASS. Systemic velocities are the
  values derived from our fitting of central data points (see Section
  \ref{results}).}
\label{gal_data}
\end{centering}
\end{table*}

\subsection{Data reduction}
\label{dr}
Data reduction were carried out using the DEIMOS DEEP2 pipeline
software. This package produces a variety of outputs from the raw
DEIMOS data, including both object and 'sky' spectra. The data
reported here probe the kinematics of the stellar populations in
the galaxy halos. These are based on the `sky' spectra that were
subtracted from the globular cluster spectra (the prime targets of the
observation, see Romanowsky et al. 2008). The 'sky´ spectra actually consist of a co-addition of the
true sky and the background galaxy light. In order to distinguish them
from the \emph{true} sky spectrum we shall therefore henceforth refer
to these as the `background' spectra.

The recovery of the galaxy spectra requires an estimate of the true
sky spectrum in each of the background spectra. For each mask, the sky
spectra were estimated from the average of four to six of the
background spectra sampling large distances from the galaxy centre
(see below). These generally lay beyond $\sim$6--7~R$_{\rm eff}$. Even
at this distance some galaxy light is still present. However, since
the data reported here are limited to within $\sim$3~R$_{\rm eff}$,
the galaxy light contamination of the sky estimates constitutes only a small
fraction of the galaxy light in our science spectra. Assuming a
de~Vaucouleur's (1953) R$^{1/4}$ profile, we estimate that, in points
at 3~R$_{\rm eff}$, the sky estimate contains only $<$10\% of the
galaxy light in the science spectra. This falls to $\sim$4\% at 2~R$_{\rm eff}$. 
Errors caused by the residual galaxy light in our sky
estimates are therefore likely to be much smaller than the quoted
errors on data points at these radii, which are often as much as $\sim$30\% of the
measured parameter values.

\begin{figure}
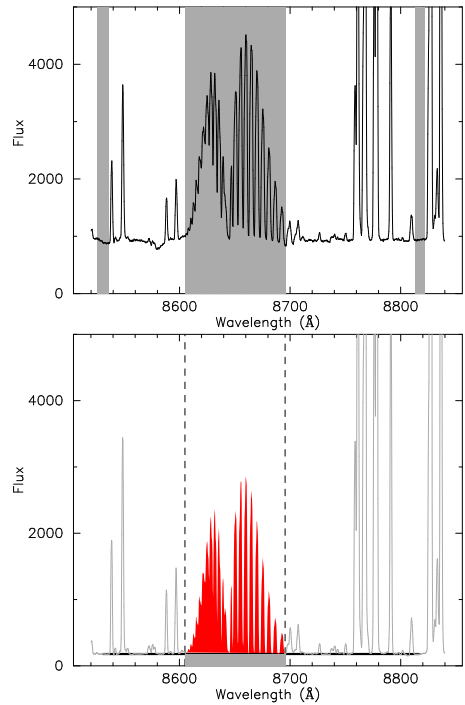

\centerline{\psfig{figure=sky_defn.ps,width=6cm,angle=-90}}
\centerline{\psfig{figure=index_defn.ps,width=6cm,angle=-90}}
\caption{Definition of the sky index. Upper plot shows wavelength
  regions of the two sidebands and the central band in a spectrum with
  significant galaxy halo signal.  Lower plot shows the region used in the
  calculation of the sky index in a scaled sky spectrum. The index is
  defined as a ratio, i.e. excess flux in the central band (counts in
  shown in red) divided by continuum flux (counts shown in grey).}
\label{skydefn}
\end{figure}

\begin{figure}
\centerline{\psfig{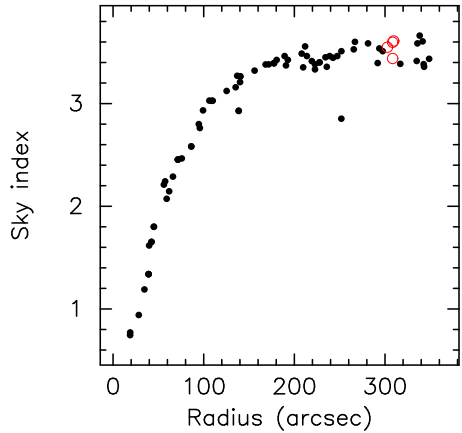}}
\caption{Sky index with radius for one set of NGC~4494 observations.
Open red circles represent points whose spectra were co-added to generate
the sky spectrum. Two points clearly exhibit aberrant behavior. These
are \emph{not} included in the final sample. }
\label{skyprofile}
\end{figure}

In order to select and characterise the sky spectra, a sky index was
defined (see Table \ref{skyind} and Fig. \ref{skydefn}). The central
band of this index was defined to cover the strong sky emission lines
between 8600 and 8700 $\AA$. The sidebands of the index were chosen to
cover regions of the background spectra that were relatively free of both
significant sky emission lines \emph{and} galaxy absorption
features. For each background spectrum the sky index was then
calculated as follows: First, the continuum level in the region of the
\emph{central} band was estimated by integrating under the linear
interpolation of the average fluxes in the sidebands.  Then the
\emph{excess} flux in the central band was calculated as the total
flux in the region minus the integrated continuum flux. Finally, the
sky index was calculated as the ratio of the excess central band flux
to its continuum flux (see Fig. \ref{skydefn}; bottom). The index has
the property that, as one moves away from the galaxy centre (and the
galaxy light contribution to the background spectrum
decreases), the index tends to a constant value (see Fig.
\ref{skyprofile}). After a careful visual inspection of the spectra,
the background spectra to be used as sky estimates
were selected from this constant region.  These were co-added to provide
a higher signal-to-noise estimate of the sky spectrum.

The index was used to select candidate sky spectra, rather than simply
choosing the slits at the greatest distance from the galaxy centre, as
the index allows identification of spectra suffering from reduction
problems not detected by visual inspection. For the most part, these
proved to be either slits located near the edges of the mask where
vignetting effects are large, or slits in which residual globular
cluster light was still present. Fig. \ref{skyprofile} illustrates how
the sky index allows identification (and elimination) of slits with
such problems.

\begin{figure}
\centerline{\psfig{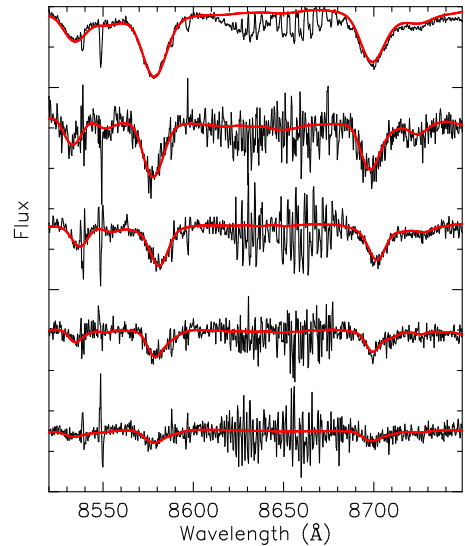}}
\caption{Sky subtracted spectra of NGC~2768. Red lines show the
best-fit template derived by the pPXF code during measurement of
velocity moments. Spectra are in order of decreasing signal-to-noise
broadly covering the range within our data (from $\sim$40 in the top
spectrum down to 7 in the bottom spectrum).}
\label{galspec}
\end{figure}

Sky subtraction was carried out by consideration of the \emph{excess
flux} in the central band as defined above. First, for each background
spectrum, we scaled the sky estimate such that the resultant spectrum
had the same excess counts as the target spectrum. Subtraction of the
scaled sky spectrum therefore resulted in a spectrum with no excess
counts in the sky band region, i.e. with counts in the sky band region
equal to the continuum level in the side bands. This method provides a
sky subtraction accurate to approximately one count over the sky band
region.  We show examples of sky subtracted spectra in
Fig. \ref{galspec} in which it can be seen that although sky residuals
are present, these are not significantly larger than the Poisson noise
associated with the strong sky-lines. Assuming a dark-sky
brightness at Mauna Kea of I $\sim$ 20 mag. arcsec$^{-2}$, then
our galaxy halo spectra can reach as low as 1--10\% of the sky background. 
We note that this sky subtraction technique is similar to that used in Proctor et
al. (2008) and Norris et al. (2008) in the analysis of optical spectra
from GEMINI/GMOS.

\begin{table}
\begin{centering}
\begin{tabular}{|c|c|}  
\hline
Band         &   Wavelengths  \\
\hline
Sideband 1    & 8526.0 -- 8536.0~$\AA$\\
Central band & 8605.0 -- 8695.5~$\AA$\\ 
 Sideband 2   & 8813.0 -- 8822.0 $\AA$\\
\hline
\end{tabular}
\caption{Band definitions for the sky index used in the selection and subtraction of sky spectra.}
\label{skyind}
\end{centering}
\end{table}

\section{Analysis Method}
\label{anal}
The analysis presented here involves measurement of velocity moments
(V, $\sigma$ and the Gauss-Hermite coefficients h$_3$ and h$_4$) using
the pPXF code of Cappellari \& Emsellem (2004) - an extension of the
pixel fitting code of van der Marel (1994).

\subsection{Measurement of kinematics}
The pPXF software works in pixel space, finding the combination of template
stars which, when convolved with an appropriate line-of-sight velocity
distribution, best reproduces the galaxy spectrum. 
Thirteen template stars were used, ranging in spectral type from F8 to M1. 
This method of
kinematic analysis has the advantage that template-mismatch
issues are all but eliminated.  Operating in pixel space, the code
also permits the masking out of regions of the galaxy spectrum during
the measurement. This is of particular importance in our data as,
despite the accurate sky subtraction, the spectra still suffer from
skyline residuals (see Fig. \ref{galspec}). The code also allows
suppression of the higher order moments h$_3$ and h$_4$ (i.e. fitting
only V$_{\rm obs}$ and $\sigma$). This was found to be useful in order
to stabilise the fits in low signal-to-noise spectra.

Errors were estimated from a repeat observation on different nights of
one of the masks of NGC~2768.  By binning the 60 repeat measurements
by observed signal (i.e. counts), the rms variations of each of the
velocity moments in the repeat measurements were calculated as a
function of signal-to-noise. The
trend of rms with signal-to-noise so obtained was then used to make
estimates of errors in this, and other galaxies.

\begin{figure}
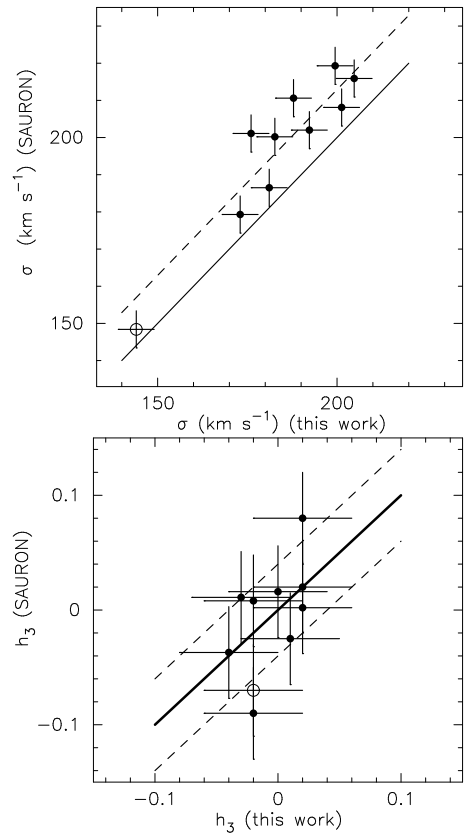

\centerline{\psfig{figure=cf_sauron_sig.ps
,width=6cm,angle=-90}}
\centerline{\psfig{figure=h3.ps,width=6cm,angle=-90}}
\caption{Comparison of velocity dispersion measures (top) and h$_{\rm 3}$ 
(bottom) with SAURON data (Emsellem et al. 2004). Filled symbols are from NGC~2768, and
the open symbol from NGC~821.  In the plot of $\sigma$, the one-to-one (solid) and
--12.9 km s$^{-1}$ offset (dashed) lines are
shown. In the plot of h$_{\rm 3}$ the one-to-one line (solid) and typical
errors (dashed) of $\pm$0.04 are shown.}
\label{cf_sauron}
\end{figure}

As a test of the accuracy of our measurements we compared them to
those of SAURON (Emsellem et al. 2004) for the two galaxies in common
between the studies. This was possible as our data samples 10 precise
locations (9 in NGC~2768 and one in NGC~821) that are also sampled by
SAURON. The results of the comparison for $\sigma$ and h$_{\rm 3}$ are
shown in Fig. \ref{cf_sauron}. A small systematic offset of --12.9 km s$^{-1}$ 
is evident in our velocity dispersion data compared to the
SAURON values, with a scatter of 7.3 km s$^{-1}$.  The measured
h$_{\rm 3}$ values are consistent within their errors.  We do not
compare h$_{\rm 4}$, as errors are generally large with respect to our
measured values. We leave comparison of V$_{\rm obs}$ to the sections
presenting our results.

During the analysis significant discrepancies were detected between
our results for the $\sigma$ values of NGC~1400 and NGC~1407 and
the long-slit data of Spolaor et al. (2008a). As a consequence, the data of
Spolaor et al. (2008a) were re-processed through the pPXF software using a
large number of template stars, resulting in significantly improved
measurements (see Appendix A).\\

\begin{table*}
\begin{centering}
\begin{tabular}{|c|c|c|c|c|c|c|c|c|c|}  
\hline
 R.A.& Dec.& V$_{\rm obs}$ & V$_{\rm obs}$ error & $\sigma$ & $\sigma$ error& h$_3$& h$_3$ error& h$_4$& h$_4$ error \\
\hline
NGC~2768&&&&&&&&&\\
   9:11:31.97 &  +60:1:38.7&  1300 &   26 &152&   24 &   0.14 & 0.04 &  0.10 & 0.05 \\
   9:11:28.47 &  +60:2:51.5&  1251 &   27 &177&   25 &   0.05 & 0.04 &  0.03 & 0.05 \\
   9:11:35.29 &  +60:2:03.8&  1284 &    5 &192&   4  &   0.02 & 0.04 &  0.02 & 0.05 \\
   9:11:38.70 &  +60:2:57.4&  1346 &   26 &154&   24 &  -0.02 & 0.04 & -0.01 & 0.05 \\
   .......... &  .........&  ...... &   ...  &.....&   .... &  ..... & .... & ..... & .... \\

\hline
\end{tabular}
\caption{Example of the velocity moment data form this work. Columns
are; R.A. and Dec (J2000), observed recession velocity (V$_{\rm obs}$
in km s$^{-1}$), velocity dispersion ($\sigma$ in km s$^{-1}$) and
Gauss-Hermite coefficients h$_3$ and h$_4$. The full table is
available on-line.}
\label{kindat}
\end{centering}
\end{table*}

\begin{figure}
\centerline{\psfig{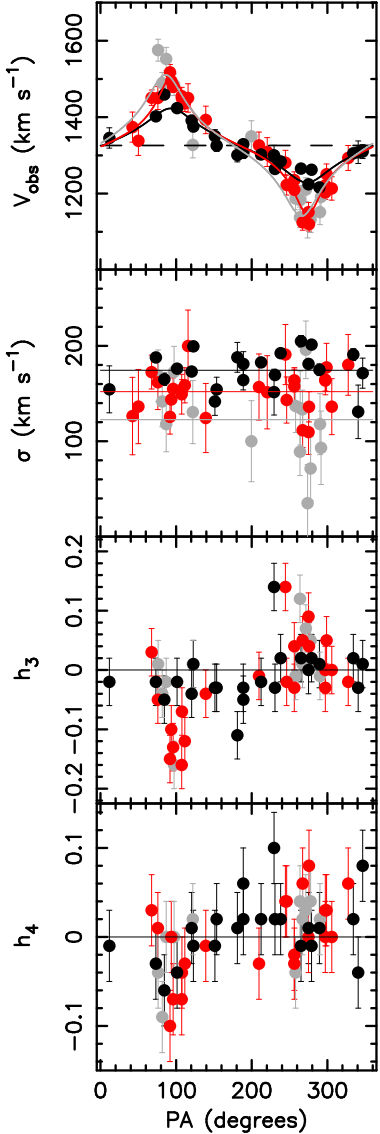}}
\caption{{\bf NGC~2768}. Velocity moments are plotted against position
angle. Data points are coloured according to galacto-centric radius
(black; R$<$1~R$_{\rm eff}$, red; 1~R$_{\rm eff}<$R$<$2~R$_{\rm eff}$
and grey; R$>$2~R$_{\rm eff}$). In the plot of V$_{\rm obs}$ best-fit 
elliptical models at each radius are shown as appropriately
coloured lines. The systemic velocity (V$_{\rm sys}$) is shown as a
dashed line. The major kinematic axis (PA$_{\rm kin}$) is at
$\sim$90\degrees. In the plot of $\sigma$ average values are also shown
as appropriately coloured lines. The values adopted for these lines
are presented in Table \ref{fits}. 
}
\label{2768theta}
\end{figure}

The measured velocity moments of all five galaxies are available on-line. An 
example of the available data are presented in Table \ref{kindat}.

\subsection{Kinemetry}
Our analysis of these data proceeds by consideration of the velocity
moments in the space of galacto-centric radius and position angle (PA;
defined as the angle between the point in the galaxy and the galaxy
centre from North through East). For the analysis, the data were
binned in radial ranges (i.e. circular shells) for comparison to the
expectations for rotating isotropic ellipsoids (or, equivalently,
inclined discs). Fits of the observed velocity (V$_{\rm obs}$) data were
carried out by least-squares minimisation of the data to the equation:

\begin{equation}
V_{\rm obs}  =  V_{\rm sys}+V_{\rm rot}.cos(\phi),
\end{equation}

where V$_{\rm sys}$ is the systemic velocity of the system, V$_{\rm rot}$ 
is the major-axis rotation velocity and $\phi$ is defined by;

\begin{equation}
tan(\phi) = \frac{tan(PA-PA_{\rm kin})}{q},
\end{equation}

with PA$_{\rm kin}$ the position angle of the major rotation axis and
q the rotation axis-ratio.  The values of V$_{\rm rot}$, PA$_{\rm kin}$ 
and q were all allowed to vary freely.  Values of V$_{\rm sys}$
were taken as the value derived from the high-signal-to-noise data of
our innermost points (i.e. generally the points within 1~R$_{\rm eff}$). 
This value was then assumed for all radial bins.

In the analysis of the rotation properties of our sample galaxies we
used two types of radial binning. The first was to break the data into
2 or 3 independent radial slices (i.e. into circular annuli), the
other was to carry out rolling fits again with data binned in circular
annuli.  Results are presented as the major axis values with radii
expressed in terms of the effective radius (R$_{\rm eff}$)\footnote{Note 
that R$_{\rm eff}$ is expressed as the radius of a circle with the
same area as the ellipse encompassing 50\% of the galaxy light. Therefore, 
in galaxies with non-zero ellipticity, a point on the major axis at
1~R$_{\rm eff}$ does \emph{not} lie on, but rather within, the
half-light ellipse.}. Errors on these fits were estimated as the rms
scatter in the results of 50 Monte Carlo simulations of the data using
the best-fit parameter values and scatter.

We were not however able to carry out such an analysis on the velocity
dispersion data, mainly because we detect little to no variation in
this parameter over the radii probed by our data. We are therefore not
able to measure the axis ratio of this velocity moment. However, for two
galaxies (NGC~821 and NGC~2768), the SAURON data of Emsellem et al. 2004 permit estimation of
axis ratios of $\sigma$ in the inner regions. For NGC~821 the SAURON
data appear to show an axis ratio close to 1.0 (despite the
ellipticity of the galaxy isophotes). On the other hand, in NGC~2768
$\sigma$ appears to have a similar axis ratio to the isophotes. We
therefore adopt these axis ratios for the analysis of the velocity dispersion 
data from these galaxies. For the remaining three galaxies (all of which have
axis ratios $\geq$0.8 we assume an axis ratio of 1.0.

We also experimented with the co-adding of spectra at similar
effective radii for $\sigma$ measurement. However, this process
requires the spectra to be de-redshifted prior to co-addition, which
complicates the handling of the sky-line residuals in the absorption
features. In addition the results proved almost identical to the
values obtained by averaging individual measurements. On the basis
that simplest is best, we therefore present $\sigma$ values as
individual and radially averaged values.

\begin{figure*}
\centerline{\psfig{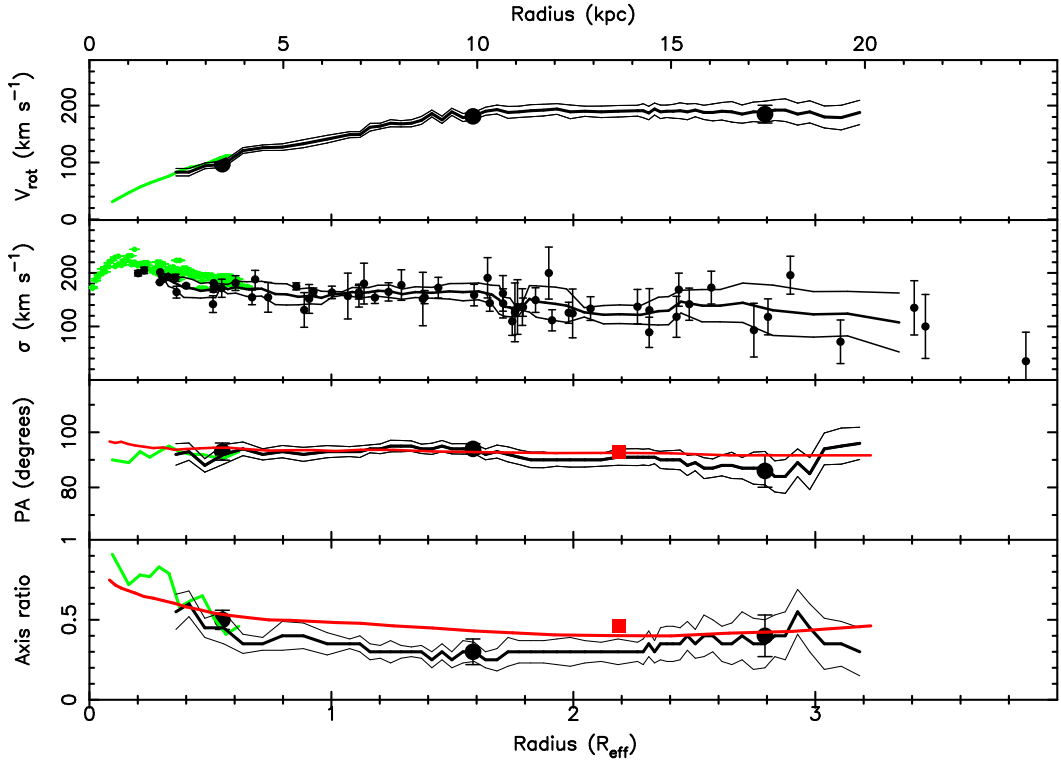}}
\caption{{\bf NGC~2768}. Radial profiles of the major axis rotation
velocity (V$_{\rm rot}$), $\sigma$, kinematic and photometric PA
(PA$_{\rm kin}$ and PA$_{\rm phot}$) and axis ratio (q and
1-$\epsilon$). Note that ellipticity of $\sigma$ is assumed to be 0.54, i.e. 
the photometric ellipticity.  The physical radial distances
(based on data in Table \ref{gal_data}) are given in kiloparsec on the
top axis.  The $\sigma$ data points (black) are the individual
measured values, with the line representing a rolling average of 5
points.  For all other parameters the black data points represent the
values of our fits to independent regions (Table \ref{fits}).  The
results of rolling fits to the data and their 1-$\sigma$ errors are
indicated as black lines. Green data points and lines at small radii
are from our analysis of the SAURON data of Emsellem et al. 2004.  The red squares in PA and
axis ratio are the 2MASS photometric values at the radius of the K
band 20th magnitude isophote, while the red lines show the R band
photometric data of Peletier et al. (1990). Overall the agreement with
the literature values is very good.  }
\label{2768rad}
\end{figure*}

Our analysis is similar in intent to the `kinemetry' of the SAURON
paper by Krajnovi{\'c} et al. (2006). However, our data possess an
extremely low sampling density (i.e we have on average 60 points
inside an area of 24~R$_{\rm eff}^2$ compared to SAURON's $\sim$3000
points inside 0.4~R$_{\rm eff}^2$ - a data density contrast of
3000:1). Our data are also generally of lower signal-to-noise than the
SAURON data.  Since, as noted in Krajnovi{\'c} et al. (2006), such
limitations make it impossible to perform a full harmonic analysis, we
have restricted our analysis to only the low order harmonics and fitted
our data on circles (or more precisely on annuli) rather than
ellipses.  Unable to perform the full analysis, we have also refrained
from adopting the full notation used by Krajnovi{\'c} et
al. (2006). However we note that the rotation velocity (V$_{\rm rot}$)
we report is most directly related to their k$_1$ parameter, while the
kinematic axis ratio is most directly related to their q$_{\rm kin}$
parameter.  PA$_{\rm kin}$ has the same meaning in both studies.

\section{Results}
\label{results}
\begin{table*}
\begin{centering}
\begin{tabular}{|c|c|c|c|c|c|c|c|}  
\hline
Galaxy& No & $<$R$>$    &$<$R/R$_{\rm eff}$$>$  &$<$$\sigma$$>$    &  V$_{\rm rot}$ & axis ratio  &  PA$_{\rm kin}$   \\
&   &(arcsec)&(R$_{\rm eff}$) &(km s$^{-1}$)&(km s$^{-1}$)&   (q)       & (\degrees)     \\
\hline			   			     		   		     
NGC~2768&21 & 31   & 0.48 &174$\pm$4      & 98$\pm$3   &   0.45$\pm$0.06   &  90$\pm3$   \\
&26 & 93   & 1.45 &152$\pm$4      &178$\pm$6   &   0.30$\pm$0.08   &  94$\pm2$   \\
&15 & 165  & 2.57 &123$\pm$10     &190$\pm$16  &   0.35$\pm$0.13   &  88$\pm6$   \\
\hline	      	    	   			    		     				   		   	    	   		   	    		     			   	   
NGC~4494&19 &  32  & 0.65 &133$\pm$3      & 58$\pm$2    & 0.80$\pm$0.03   & 182$\pm$1 \\
&30 &  75  & 1.53 &111$\pm$3      & 52$\pm$3    & 0.50$\pm$0.09   & 179$\pm$3 \\
&33 & 143  & 2.91 &100$\pm$5      & 54$\pm$5    & 0.40$\pm$0.20   & 170$\pm$6 \\
\hline	      	    	   			    		     				   		   
NGC~1407&28 &  49  & 0.70 &261$\pm$6      & 28$\pm$4   & (0.95)      & 254$\pm$8\\
&39 &  102 & 1.46 &240$\pm$7      & 21$\pm$6   & (0.95)      & 245$\pm$14\\
&14 &  173 & 2.48 &198$\pm$11     & 20$\pm$10  & (0.95)      & 248$\pm$15\\
\hline	      	    	   			    		     				   		   
NGC~1400& 7 & 31   & 1.07  &116$\pm$3      & 53$\pm$3     &(0.9)       & 34$\pm$4\\
&14 & 68   & 2.35  &96$\pm$7       & 30$\pm$7     &(0.9)       & 40$\pm$14\\
\hline	      	    	   			     	     				   		   
NGC~821&14 &   41 & 0.82 & 149$\pm$5     & 56$\pm$8   &   0.30        &   33  \\
&17 &   78 & 1.56 & 144$\pm$6     &  6$\pm$8   &  (0.30)       &  (33)  \\
\hline
\end{tabular}
\caption{Kinematic data summary for independent radial slices.  The
definition of the radial binning is outlined in Section \ref{results}.
The number of slits in each radial bin, their average radius and
$\sigma$ are given as well as the results of the rotation velocity
fits (Equations 1 \& 2). Data in brackets are assumed (rather than
fitted) values.}
\label{fits}
\end{centering}
\end{table*}

\begin{figure*}
\centerline{\psfig{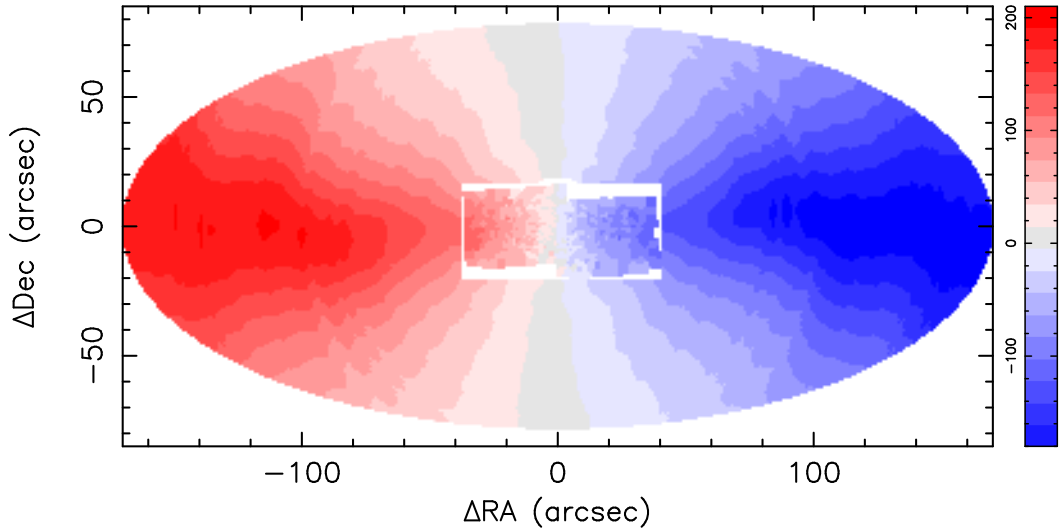}}
\caption{{\bf NGC~2768}.  Reconstruction of the 2-D rotation velocity
map using the rolling fits of Fig.
\ref{2768rad}. The outer boundary of the map has been chosen to
reflect the photometric axis ratio of the galaxy
(0.46). The colour scale (in km s$^{-1}$) is shown on the right.
Also shown is the SAURON (Emsellem et al. 2004) velocity map of the central regions
of the galaxy. The agreement between the data sets is extremely good.}
\label{2768image}
\end{figure*}

In this section we present the results of our velocity moment
measurements (V$_{\rm obs}$, $\sigma$, h$_3$ and h$_4$).  Errors on
individual data points were derived from the repeat measurement of one
mask in NGC~2768 (see Section \ref{anal}). Errors on rolling fits were
estimated by Monte Carlo realisations of the best-fit parameters.
Note that, for the lowest signal-to-noise spectra, the higher order
moments were either suppressed altogether or are not presented. Tests
on the low signal to noise spectra show that changes in recession
velocity and velocity dispersion caused by the suppression of the
h$_3$ and h$_4$ parameters are $\sim$20\% of the quoted errors.
We also present re-constructions of 2-D rotation velocity maps in this
section.

The results of the fits to independent radial slices are presented in
Table \ref{fits}. In NGC~821 and NGC~1400 two radial ranges were
defined (NGC~821: R$<$1.0~R$_{\rm eff}$ and R$\geq$1.0~R$_{\rm eff}$,
NGC~1400: R$<$1.5~R$_{\rm eff}$ and R$\geq$1.5~R$_{\rm eff}$).  In the
other three galaxies three ranges were defined (R$<$1.0~R$_{\rm eff}$,
1.0~R$_{\rm eff}\leq$R$<$2.0~R$_{\rm eff}$ and R$\geq$2.0~R$_{\rm eff}$). 

We first present our results for
NGC~2768, as this galaxy possesses high signal-to-noise data, a large
sample size, and is also one of the galaxies observed by SAURON.

\subsection{NGC~2768}
\label{NGC2768}

The brightest galaxy in a loose group, NGC~2768 is classified as an E6
galaxy in RC3 (de Vaucouleurs et al. 1991). However, in both the
Revised Shapley--Ames Catalogue of Bright Galaxies (Sandage, Tammann
\& van den Bergh 1981) and the Carnegie Atlas of Galaxies (Sandage \&
Bedke 1994) it is classified as an S0$_{1/2}$ due to a `definite outer
envelope'. The galaxy is highly elliptical with K band photometric
axis ratio of 0.46, and PA of 93\degrees (Jarrett et al. 2000).  It
also reveals a polar orbiting dust and gas ring, possibly of external
accretion origin (see Martel et al. 2004 and references therein).

Photometric profiles out to 200 arcsec are presented in Peletier et
al. (1990). They find a rapidly rising ellipticity profile with radius
(i.e. a falling axis ratio), and report indications of a `disk-like component'
at about 50 arcsec which gives way to boxy isophotes beyond about 100
arcsec.

Fried \& Illingworth (1994) probed the kinematics along the minor and
major axes out to about 50 arcsec.  They found evidence for stellar
rotation aligned along the photometric major axis and a `spheroidal'
velocity field.  They also detected minor axis rotation associated
with the polar ring.

The central regions ($\sim$80x40 arcsec) of the galaxy were also
observed with the SAURON integral field unit (Emsellem et
al. 2004). The SAURON results show strong rotation with a `cylindrical
velocity field' and a central dip in $\sigma$ (McDermid et
al. 2006). They find the galaxy to be a fast-rotator, but, contrary to
Sandage \& Bedke (1994), conclude that the galaxy `...does not have
evidence for a disc'. They therefore adopt the RC3 classification of
this galaxy as an E6.
 
The SAURON team also investigated the very central regions ($\sim$ 4x4
arcsec) using the OASIS integral field spectrograph (McDermid et
al. 2006), identifying emission associated with the polar disk and a
young (2.5~Gyr) central stellar population. However, other
spectroscopic age determinations in the literature generally prove
inconsistent, with studies reporting the central stellar population to
be both old and young (e.g. Howell 2005; Denicolo et al. 2005;
Sil'Chenko 2006). It is unclear whether this reflects inconsistencies in
the determination of the properties of the central populations, or is rather a real
effect, perhaps caused by variable sampling of the central
star-forming disk.

Hakobyan et al. (2008) suggest this galaxy may be a merger remnant
with the presence of a young stellar population suggested by ``{\it...the
existence of strong HI and CO emission, and the presence of dust and
ionised gas..}''. The recent type Ib supernova SN2000ds (Filippenko \&
Chornock 2000) supports this conclusion.\\

Presenting our results, we first consider the distribution of the
NGC~2768 velocity moments with PA.  Results are plotted in
Fig. \ref{2768theta} with point colour depicting galacto-centric
radius. Values derived for each radial bin are presented in Table
\ref{fits}.\\ 
\indent The rotation in this galaxy can be seen to be strongly
elliptical. An increasing rotation velocity with radius is also
evident.  The h$_3$ parameter reflects this behaviour, with near-zero
values in the slowly rotating inner regions (R$<$1~R$_{\rm eff}$). 
However, at larger radii h$_3$ increasingly mirrors the
elliptical rotation curve.  The increasing signal in h$_3$ with radius
is also detected in the SAURON maps of the galaxy (Emsellem et
al. 2004).

Both $\sigma$ and h$_4$, however, exhibit unusual behaviour, as they
do not possess the minor-axis symmetry expected from even moments (see
Krajnovi\'{c} et al. 2008). This must be understood in terms of the
degeneracy between $\sigma$ and h$_4$ discussed at length in
Cappellari \& Emsellem (2004). The degeneracy is worst when; i)
signal-to-noise is low, and ii) when two distinct kinematic components
have roughly equal luminosities. To test that this was indeed causing
the lack of axi-symmetry in our data, we re-fitted the data for this
galaxy using pPXF with h$_3$ and h$_4$ suppressed (i.e. set to
zero). The test confirmed that, if h$_4$ is forced to zero, the dip in
$\sigma$ becomes visible on both sides of the galaxy.\\ 
\indent Given that the h$_4$ parameter is more prone to the effects of
low signal-to-noise and sky-line residuals than $\sigma$, we conclude
that, in its outer reaches, this galaxy probably exhibits a
significant drop in $\sigma$ on both sides of the major axis, as would
be expected from the disc of an S0 galaxy.

Velocity dispersion appears to be decreasing with increasing radius
(Fig. \ref{2768theta} and Table \ref{fits}).  However, the dip on the
major axis (PA$\sim$270\degrees) is clearly effecting measured average
values, particularly since the slits in the outer regions of the
galaxy from which we were able to recover a galaxy spectrum tend to
lie along the major axis where the galaxy light is brightest
(Fig. \ref{2768theta}). The gradient in $\sigma$ presented here may
therefore be somewhat exaggerated by this effect. We note for future
reference that the four data points with $\sigma<$100 km s$^{-1}$ all
lie very close to the major rotation velocity axis.

To measure variations with radius we carried out a rolling fit to the
recession velocity data in annuli at increasing radii allowing V$_{\rm rot}$,
PA$_{\rm kin}$ and kinematic axis ratio (q in Equation 2) to vary with
radius. The fits were performed using a minimum of 10 data points,
increasing to 20 in the outer regions.  Results of these fits are
presented in Fig. \ref{2768rad}. We also performed rolling fits to the
SAURON data (Emsellem et al. 2004) for comparison.

The agreement in V$_{\rm rot}$ between our results and the SAURON
data is good, with near identical rotation velocities over the whole
$\sim$20 arcsec overlap region. However, the rotation velocity 
clearly continues to rise beyond the extent of the SAURON data,
with the rotation velocity doubling to $\sim$200 km s$^{-1}$ between 1
and 3~R$_{\rm eff}$.  This has a significant impact on the value of
V/$\sigma$ obtained for this galaxy, underlining the importance of
data at large radii in the construction of even simple kinematic
descriptions of galaxies (see Section \ref{discuss}).

\begin{figure}
\centerline{\psfig{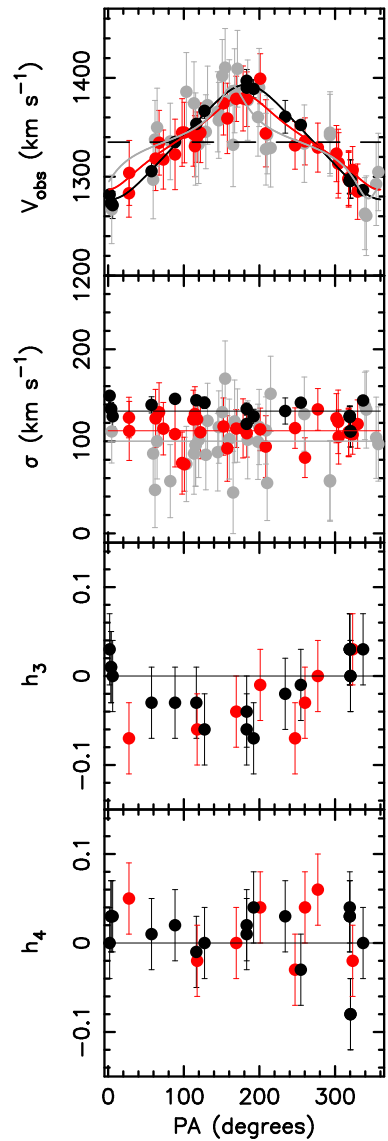}}
\caption{{\bf NGC~4494}. Same as Fig. \ref{2768theta}. The major
axis PA is 173\degrees}
\label{4494theta}
\end{figure}

\begin{figure*}
\centerline{\psfig{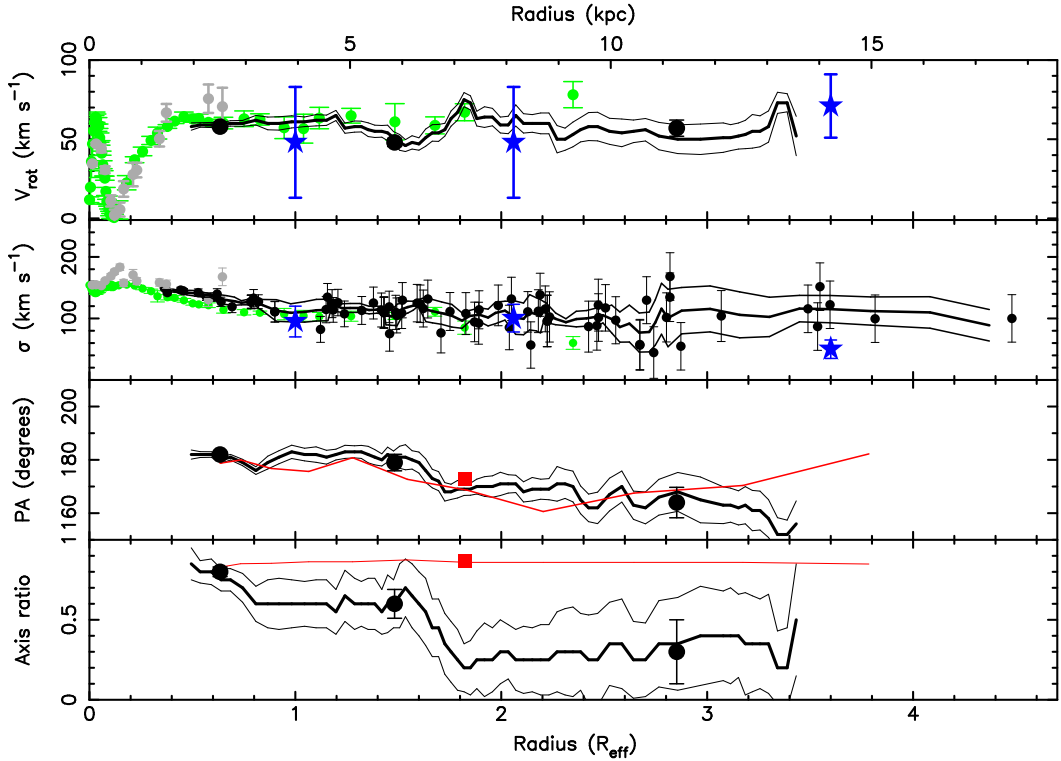}}
\caption{{\bf NGC~4494}. Same as Fig. \ref{2768rad} with the following
exceptions. Green data points in V$_{\rm rot}$ and $\sigma$ plots are
the major axis data from Napolitano et al. (2009). The planetary
nebula data (blue stars) are also from Napolitano et al. (2009).
Grey data points are the data of Bender, Saglia \& Gebhardt (1994).
Red lines in plots of PA and axis ratio are the V band results of
Napolitano et al. (2009). Large offsets between the kinematic and
photometric axis ratio and position angles can be seen.}
\label{4494rad}
\end{figure*}

Also shown in Fig. \ref{2768rad} is a plot of $\sigma$ with radius.
In this plot the agreement between the SAURON results (Emsellem et
al. 2004) and ours can be seen to be reasonable.  However, the
comparison of $\sigma$ presented in Fig. \ref{cf_sauron} is largely
based on data from this galaxy, and exhibits a $\sim$13~km s$^{-1}$
systematic offset. The results indicate a strong radial trend in $\sigma$
within 1~R$_{\rm eff}$, which levels off somewhat beyond.  However, we
again note the increasing tendency for our outer data points to lie
along the major kinematic axis and the trend for such points to
possess depressed $\sigma$. To get some idea of the impact these
trends might have on our results, consider the four points with the
lowest $\sigma$ ($<$100 km s$^{-1}$), which, as noted in the previous
section, all lie very close to the major kinematic axis. If these
points are excluded from our considerations we would conclude that
$\sigma$ was flat beyond $\sim$1~R$_{\rm eff}$.

We detect a PA$_{\rm kin}$ of 91$\pm$3\degrees throughout the galaxy
in good agreement with the photometric PA from 2MASS (Jarret et
al. 2000) and Peletier et al. (1990).  We detect no variation in
PA$_{\rm kin}$ with radius, with our measurements of the SAURON data
yielding the same, near-constant value.  

Comparison of our rotation axis ratio measurements to those of SAURON
again give good agreement in the overlap region. The results clearly
indicate that the rotation in the central regions, variously described
as `cylindrical' (Emsellem et al. 2004) or `spherical' (Fried \&
Illingworth 1994), rapidly give way to much more disk-like values at
larger radii.  Interestingly, the comparison with the photometric axis
ratios of 2MASS and Peletier et al. (1990) show the rotation axis
ratio to pass from less than the photometric axis ratio in the outer
regions to greater than the photometric axis ratio in the inner
regions.  The most natural explanation for the galaxy having a
smaller photometric than kinematic axis ratio in the inner regions is
the presence of a highly inclined, but inflated disc, such as would be
found in an S0 galaxy, in accord with the Sandage et al. (1981) and
Sandage \& Bedke (1994) classifications.

To aid in the visualisation of our results, and to facilitate the
comparison of our results with those of SAURON, we have reconstructed
the 2-D velocity map of NGC~2768 in Fig. \ref{2768image}. The
image was generated using the rolling fit values of V$_{\rm rot}$,
PA$_{\rm kin}$ and rotation axis ratio of
Fig. \ref{2768rad}. Agreement with the SAURON results can be seen to
be extremely good. Also evident is the progression from `cylindrical'
rotation in the inner, bulge dominated regions to discy
rotation in the outer regions probed by our data.

In summary, our results for this galaxy are all in accord with the
Shapley--Ames Catalogue (Sandage et al. 1981) and Carnegie Atlas
of Galaxies (Sandage \& Bedke 1994) classifications of this galaxy
as an S0, with the disk-like kinematics in the outer regions giving
way to the more `cylindrical' rotation in the inner regions.

\subsection{NGC~4494}
\label{NGC4494}
NGC 4494 is an elliptical galaxy in the Coma I cloud (Forbes et
al. 1996) with a K band photometric position angle of 173\degrees and
axis ratio of 0.87 (Jarret et al. 2000). It reveals a small central
dust ring with $\sim$60 degree inclination (photometric axis ratio
$\sim$~0.5) aligned with the major axis of the galaxy (Carollo et
al. 1997). The central regions also host a kinematically distinct core
(Bender 1988). The mass modeling analysis of Kronawitter et al. (2000)
indicates a rising M/L profile. However, evidence for dark matter in
NGC 4494 has been questioned by Romanowsky et al. (2003) based on an
extended $\sigma$ profile from planetary nebulae kinematics (but see
also Dekel et al. 2005 and Napolitano et al. 2009). The stellar
population analysis of Denicol{\'o} et al. (2005) found an
intermediate age of 6.7~Gyr in the central regions, indicating that
this is not a simple, old, passively evolving elliptical galaxy.

Our analysis of this galaxy was the same as that for NGC~2768.  The
distribution of the velocity moments with PA are shown in
Fig. \ref{4494theta}. Results show the rotation in this galaxy to be
strongly elliptical, but rotation velocity shows no sign of variation
with radius. Within 1~R$_{\rm eff}$, the h$_3$ parameter shows the
anti-correlation with V$_{\rm rot}$ expected from a disc embedded in a
more slowly rotating (or non-rotating) bulge, although it is not
possible to trace this behaviour out to the outer regions where the
signal-to-noise is low. We find a decreasing $\sigma$ with radius
which flattens out beyond 1~R$_{\rm eff}$. We find no evidence for
variation of $\sigma$ with PA.

\begin{figure}
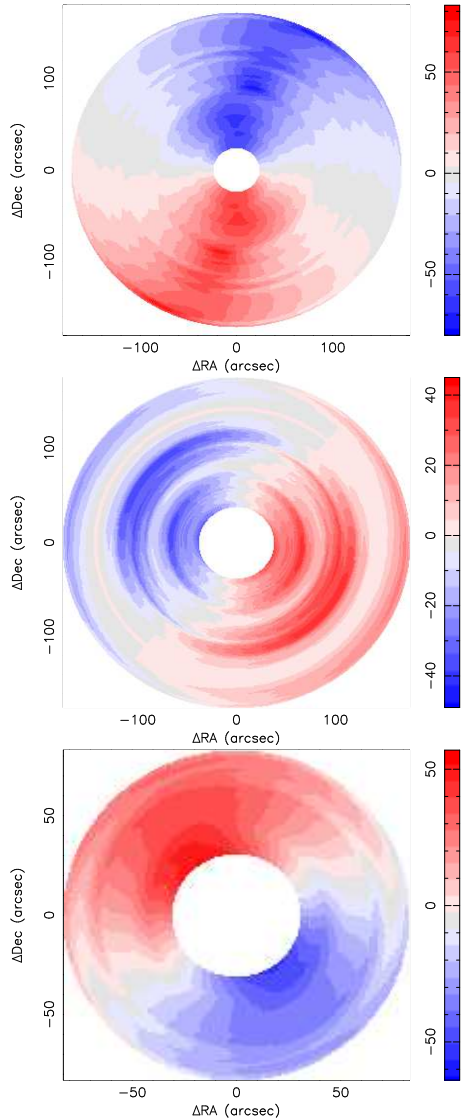

\centerline{\psfig{figure=new_image4494.ps,width=6cm,angle=-90}}
\centerline{\psfig{figure=new_image1407.ps,width=6cm,angle=-90}}
\centerline{\psfig{figure=new_image1400.ps,width=6cm,angle=-90}}
\caption{Reconstructed 2-D velocity maps of (top to bottom)
  NGC~4494, NGC~1407 and NGC~1400. Colour scales (in km s$^{\rm -1}$)
  are shown to the right of each image.  Note that the variations in
  V$_{\rm rot}$ with radius in NGC~1407 that causes the unusual
  patterning in the reconstruction are only detected at
  $\sim$1-$\sigma$ significance.}
\label{images}
\end{figure}

The rolling fits to V$_{\rm obs}$ are presented in Fig \ref{4494rad}.
These fits were performed using 10 data points in the inner region
rising to 20 data points in the outer regions of the galaxy. For
V$_{\rm rot}$, we find good agreement with profiles in the
literature. Our results are also consistent with the planetary nebulae
data of Napolitano et al. (2009) in the outer regions (notwithstanding their
large uncertainties). As deduced from
Fig. \ref{4494theta}, the galaxy exhibits no sign of variation in
rotation speed with radius over the whole radial range from
1--3~R$_{\rm eff}$. The reconstructed 2-D velocity map of this galaxy is shown in Fig. \ref{images}.

Agreements with the $\sigma$ profiles and planetary nebulae data in
the literature are again good, with $\sigma$ declining steadily with
radius within 1~R$_{\rm eff}$, but leveling off 
beyond. Indeed, beyond 1~R$_{\rm eff}$ , the data are consistent with
\emph{no} slope in $\sigma$ at the 1-$\sigma$ confidence level.

The kinematic position angle exhibits a shift of $\sim$~10~\degrees at
$\sim$1.5~R$_{\rm eff}$, mirroring the shift in photometric position
angle (Napolitano et al. 2009). 

The rotation axis ratio is similar to the photometric axis ratio in
the innermost regions, but falls rapidly at the radius at which the
position angle changes. The photometric axis ratio (Napolitano et
al. 2009), on the other hand, remains near constant with radius. The
disparity in axis ratios is all the more curious when the lack of
photometric evidence for an embedded disc at these radii (Carollo et
al. 1997) is considered. However, it is interesting to note that the
kinematic axis ratio in the outer regions of the galaxy is similar to
the photometric axis ratio of the central embedded disc of $\sim$0.5
(Carollo et al. 1997). This suggests that the inner disc could in fact
be part of a much larger (warped) disc structure.

The interpretation of the kinematics in this galaxy is clearly complex. 
The highly elliptical V$_{\rm rot}$ and h$_3$ at $\sim$1~R$_{\rm eff}$ 
and their possible association with the central disc would
seem to suggest that this galaxy harbours an embedded stellar disc at
large radii. However, the lack of photometric evidence for such a disc
raises serious questions about such an interpretation. An alternative
description is suggested by the results of Balcells (1991) who shows
that the signature of rotation accompanied by an h$_3$ signal can be
generated by a minor merger event, as a result of the re-distribution
of the angular momentum of the merging galaxy. Such an event could
also be responsible for the shift in photometric and kinematic
position angles at $\sim$1.5~R$_{\rm eff}$. Our results are therefore
at least qualitatively consistent with the idea that this galaxy is a
merger remnant. The presence of a kinematically distinct core (Bender
1988) also supports this interpretation.

\subsection{NGC~1407}
\label{NGC1407}
\begin{figure}

\centerline{\psfig{figure=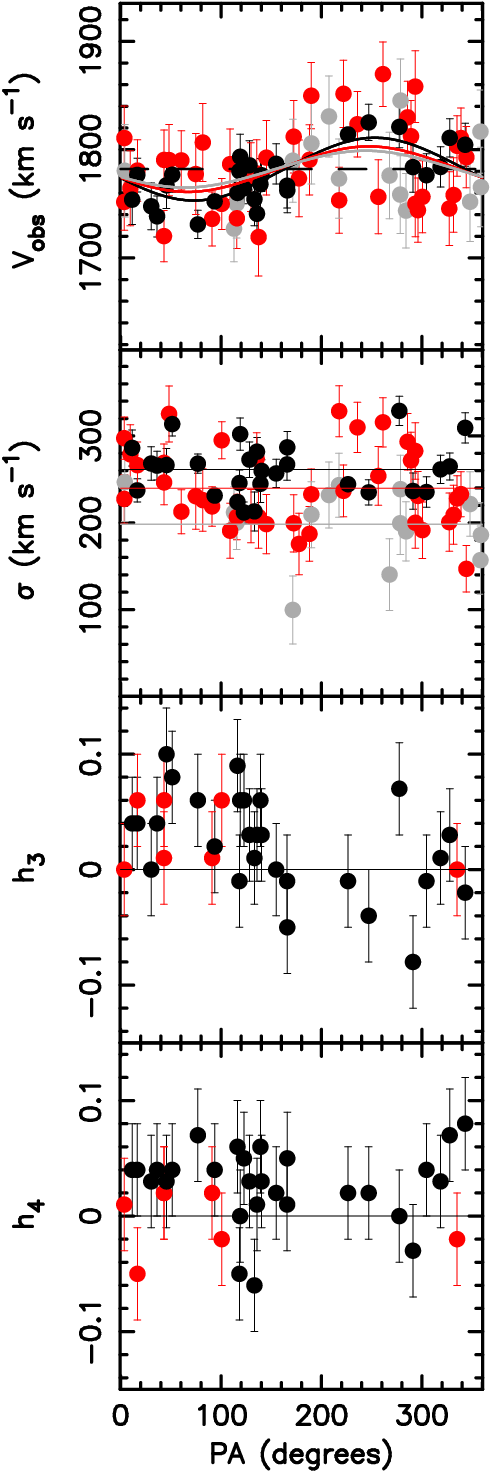,width=5cm}}
\caption{{\bf NGC~1407}. Same as Fig. \ref{2768theta} except that the
rotation axis ratio was not fitted but assumed to be equal to the 2MASS
photometric axis ratio (i.e. 0.95). The major axis PA is 60\degrees.}
\label{1407theta}
\end{figure}

NGC~1407 is the brightest group galaxy in a dwarf galaxy dominated
group (Trentham et al 2006). The galaxy has been classified as a
weakly rotating E0 (Longo et al. 1994) with a ``core-like'' central
luminosity profile (Lauer et al. 1995; Spolaor et al. 2008a).  The
stellar population analysis of Spolaor et al. (2008b) found the galaxy
to possess a uniformly old age within $\sim$0.6~R$_{\rm eff}$.

Plots of velocity moments with PA are presented in
Fig. \ref{1407theta}. The data were not fitted for kinematic axis
ratio due to low signal-to-noise and the large uncertainties in fits
of axis ratio when the value approaches 1.0. However, it is clear from
the plot of V$_{\rm obs}$ with PA that the rotation has very low
ellipticity (kinematic axis ratio $\sim$1.0), similar to the 2MASS
photometric axis ratio of 0.95. This value was therefore assumed for
the rolling fits.  A slow ($\sim$25 km s$^{-1}$) rotation is evident
in all three radial bins.  There is a suggestion of non-zero h$_3$ in
the inner regions, consistent with the results of the re-analysis of
Spolaor et al. (2008a) data (see Appendix A). The data suggest a
declining $\sigma$ with radius.

\begin{figure*}
\centerline{\psfig{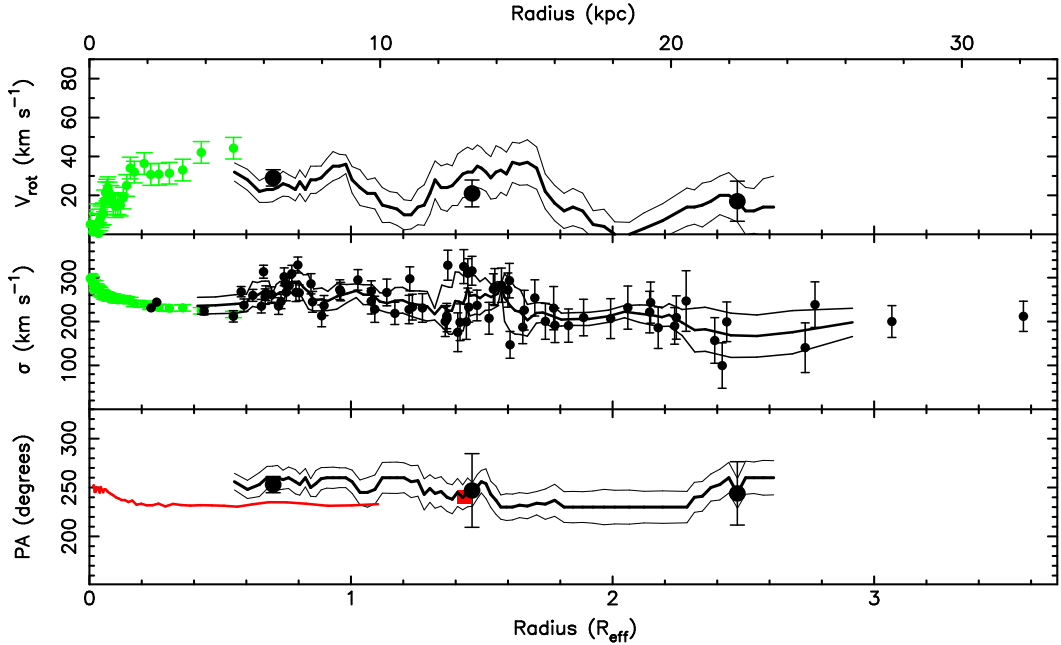}}
\caption{{\bf NGC~1407}. Same as Fig. \ref{2768theta} with the
following exceptions. Kinematic axis ratio is not presented as the
proximity of the rotation axis ratio to 1.0 makes its constraint
beyond the capabilities of the signal-to-noise of our data.  In
the V$_{\rm rot}$ and $\sigma$ plots green points are from our re-measurement
of the data of Spolaor et al. (2008a) (see Appendix A).  The
red line in the PA plot is the I band data of Goudfrooij et al.
(1994). Photometric PAs are shown with a 180 degree offset for
the purpose of comparison to the kinematic data.}
\label{1407rad}
\end{figure*}

Rolling fits were performed on the V$_{\rm obs}$ data using 12
data points in the inner region rising to 20 data points in the outer
regions of the galaxy (Fig. \ref{1407rad}).  The results confirm the
near constant V$_{\rm rot}$ with radius between $\sim$0.5 and
3~R$_{\rm eff}$ (we note that the variations with radius are only of
order 1-$\sigma$ significance).  Fig. \ref{1407rad} also confirms the shallow
decline in $\sigma$ with radius identified above. The reconstructed 2-D velocity map of this galaxy is shown in Fig. \ref{images}.

PA$_{\rm kin}$ shows a $\sim$2--$\sigma$ deviation from the
photometric PA of Goudfrooij et al. (1994) in the inner regions
($\lesssim$~1.0~R$_{\rm eff}$), but is in good agreement with the
2MASS photometric PA at 1.5~R$_{\rm eff}$. The cause of the apparent
discrepancy in the inner regions is unclear, although significant
triaxiality is one possible cause. However, it may also simply be the
result of the large uncertainties that arise when fitting both
photometric and kinematic data when axis ratios approach unity.

\subsection{NGC~1400}
\label{NGC1400}

Variously classified as an E (e.g. Da Costa et al. 1998) and an S0
(e.g. 2MASS; Jarrett et al. 2003), this galaxy is associated, but
apparently not interacting with, the NGC~1407 group. Spolaor et
al. (2008b) found the stellar population to possess a uniformly old age
out to $\sim$1.3~R$_{\rm eff}$.

\begin{figure}
\centerline{\psfig{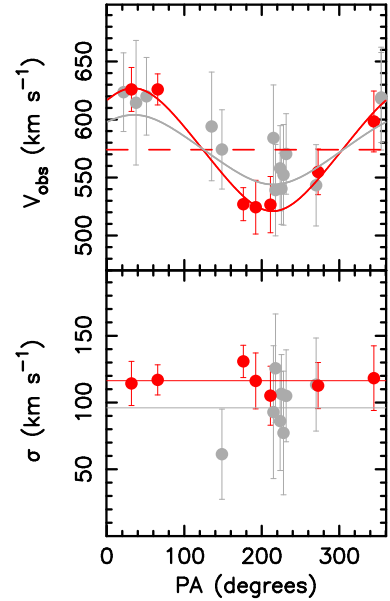}}
\caption{{\bf NGC~1400}. Same as Fig. \ref{2768theta} with the
following exceptions. The velocity moments h$_3$ and h$_4$ were suppressed
in this low signal-to-noise sample. Radial bins of the independent
data points are R$<$1.5~R$_{\rm eff}$ and R$\geq$1.5~R$_{\rm eff}$. The
rotation axis ratio was not fitted but assumed to be 0.9 (i.e. equal to
the 2MASS photometric axis ratio).}
\label{1400theta}
\end{figure}

The distribution of V$_{\rm obs}$ and $\sigma$ with PA are shown in
Fig. \ref{1400theta}. As for NGC~1407, the kinematic axis ratio was
not fitted. Instead, the 2MASS photometric axis ratio of 0.9 was
assumed. For this galaxy, the signal-to-noise was too low for the
accurate measurement of the h$_3$ and h$_4$ moments. These were were
therefore suppressed during kinematic analysis using the pPXF code
(Section \ref{anal}). However, we note the significant h$_3$ detection
in the re-analysis of the Spolaor et al. (2008a) data (Appendix
A). Our data sampled only one data point within 1~R$_{\rm  eff}$. 
Radial binning was therefore carried out in the bins
R$<$1.5~R$_{\rm eff}$ and R$\geq$1.5~R$_{\rm eff}$.  For some of the
outermost points errors in velocity dispersion exceeded
100~km~s$^{-1}$. These data were omitted from both plots and analysis.

Fig. \ref{1400theta} shows that V$_{\rm rot}$ is lower in the outer
radial range than in the inner range. There is also the suggestion of
a falling $\sigma$ with radius. The data were not fitted for rotation
axis ratio due to low signal-to-noise and the large uncertainties in
fits in axis ratio measurements when the value approaches
1.0. However, it is clear from the plot of V$_{\rm obs}$ with PA that
the rotation has very low ellipticity (rotation axis ratio $\sim$1.0),
similar to the 2MASS photometric axis ratio of 0.90. This value was
therefore assumed for the rolling fits.

\begin{figure*}
\centerline{\psfig{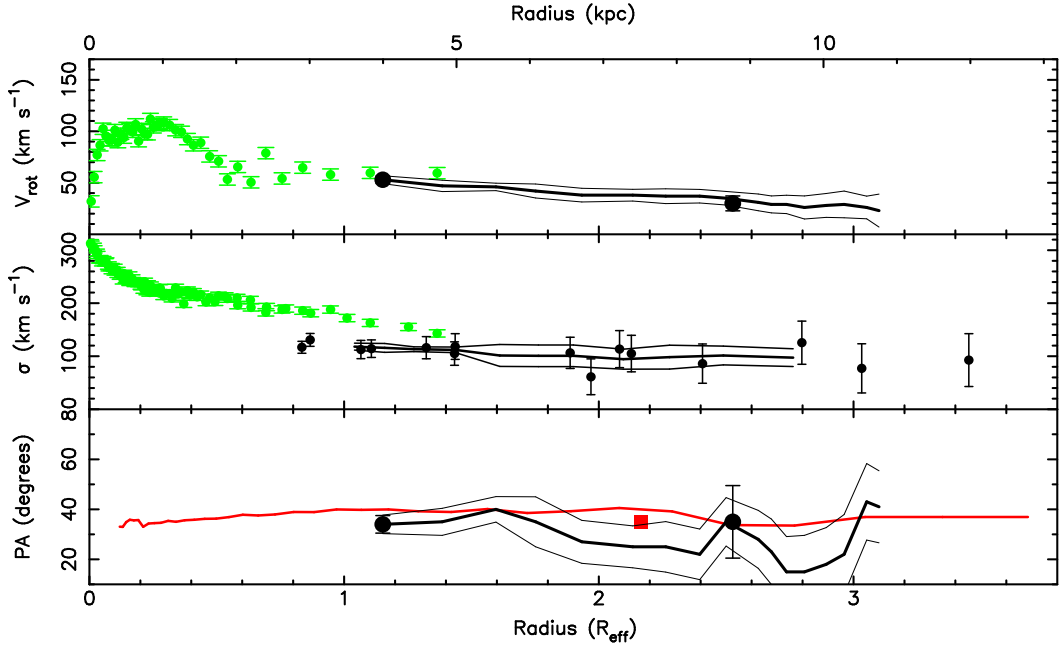}}
\caption{{\bf NGC~1400}. Same as Fig. \ref{2768rad} with the
following exceptions. The independent data points are for R$<$1.5~R$_{\rm
eff}$ and R$\geq$1.5~R$_{\rm eff}$. Green data points in V$_{\rm rot}$ and
$\sigma$ plots are from the re-measurement of the data of Spolaor et
al. (2008a) (see Appendix A). The red line in the PA plot is
the Sloan i band data of Spitler et al. (2009; in prep).}
\label{1400rad}
\end{figure*}

Due to the small sample size, rolling fits were performed to V$_{\rm obs}$ 
using 7 points in the inner region rising to 12 in the outer
regions.  The resultant plots with radius (Fig. \ref{1400rad}) confirm
the conclusions drawn from the plot with PA of a falling V$_{\rm rot}$
and $\sigma$ with radius. Indeed, this galaxy shows the sharpest
fall-off in $\sigma$ with radius in our sample - falling from
$>$300~km s$^{-1}$ in the centre to $\sim$100~km s$^{-1}$ at
$\sim$1.5~R$_{\rm eff}$, but remaining near constant beyond.  However,
there appears to be a significant unexplained offset between $\sigma$
from this work and the values from the data of Spolaor et al. (2008a)
in the overlap region (both before \emph{and} after the re-measurement
of the data using the pPXF; See Appendix A).  It is important to note
that this discrepancy increases with decreasing radius, with the
discrepancy rising from $\sim$~20~km s$^{-1}$ at 1.4~R$_{\rm eff}$ to
60 km s$^{-1}$ at 1~R$_{\rm eff}$ (see also Fig. \ref{cf_sauron}).
This clearly indicates that this is not a
signal-to-noise issue. A detailed examination of the spectra of the
two studies of NGC~1400 could identify no obvious reasons for such a
discrepancy. Consequently, we must contemplate the possibility that
this is a real effect. We therefore note that the sense of the
discrepancy is consistent with the idea that the optical and NIR probe
different stellar populations, with the NIR being dominated by a lower
velocity dispersion (more centrally concentrated) population than the
optical. The strong stellar population metallicity gradient in NGC~1400
(--0.38 dex/dex; Spolaor et al. 2008b) may therefore be the cause of the
large discrepancy in this galaxy. The reconstructed 2-D velocity map of this galaxy is shown in Fig. \ref{images}.

PA$_{\rm kin}$ shows no significant variation with radius and is in good
agreement with the photometric PA of 2MASS and Spitler et al. (2009; in prep).

With regard to this galaxy's rather uncertain morphological
classification, the evidence for a disc with a declining rotation, but
near constant $\sigma$ profile would seem to favour its classification
as an elliptical with an embedded disc.

\subsection{NGC~821}
\label{NGC821}

NGC 821 is an isolated (Reda et al. 2007), elliptical galaxy with
morphological classification E6.
The galaxy is considered to possess a disk embedded in a
spheroidal halo (Michard \& Marchal 1994; Ravindranath et al. 2002;
Emsellem et al. 2004).  The photometric profiles of Goudfrooij et
al. (1994) show that ellipticity peaks around 15 arcsec
($\sim$0.3~R$_{\rm eff}$).  Kinematic profiles have been widely
published (e.g. Pinkney et al. 2003; Proctor et al. 2005; Forestell \&
Gebhardt 2008), although there is still some debate as to whether the
$\sigma$ profile is falling or not and whether NGC 821 contains dark
matter (Romanowsky et al. 2003). The stellar population study of
Proctor et al. (2005) reveals a young central stellar population which
gives way to old ages by 1 effective radius.

\begin{figure}
\centerline{\psfig{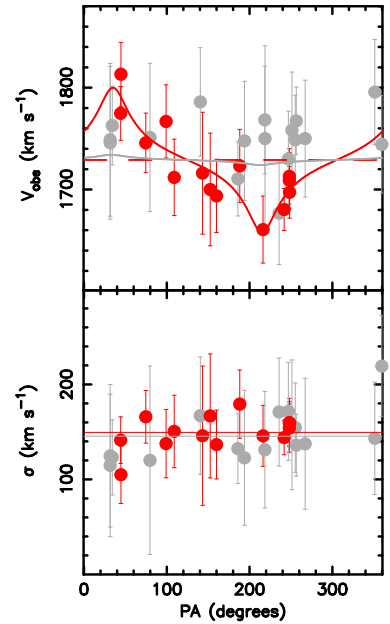}}
\caption{{\bf NGC~821}. Same as Fig. \ref{1400theta} except that the
 radial binning is in ranges of R$<$1.0~R$_{\rm eff}$ and
R$\geq$1.0~R$_{\rm eff}$ only. The major axis PA is 30\degrees.}
\label{821theta}
\end{figure}

Plots of velocity moments with PA are shown in
Fig. \ref{821theta}. Only the data within 1~R$_{\rm eff}$ were fitted
for position angle and rotation axis ratio due to low signal-to-noise
and small sample size in the outer regions. The values derived from
the inner regions was then assumed for the outer regions.

Fig. \ref{821theta} shows
strong rotation within 1~R$_{\rm eff}$, but exhibits a value
consistent with zero rotation beyond this region. We also detect no
variation in $\sigma$ with radius. 

\begin{figure*}
\centerline{\psfig{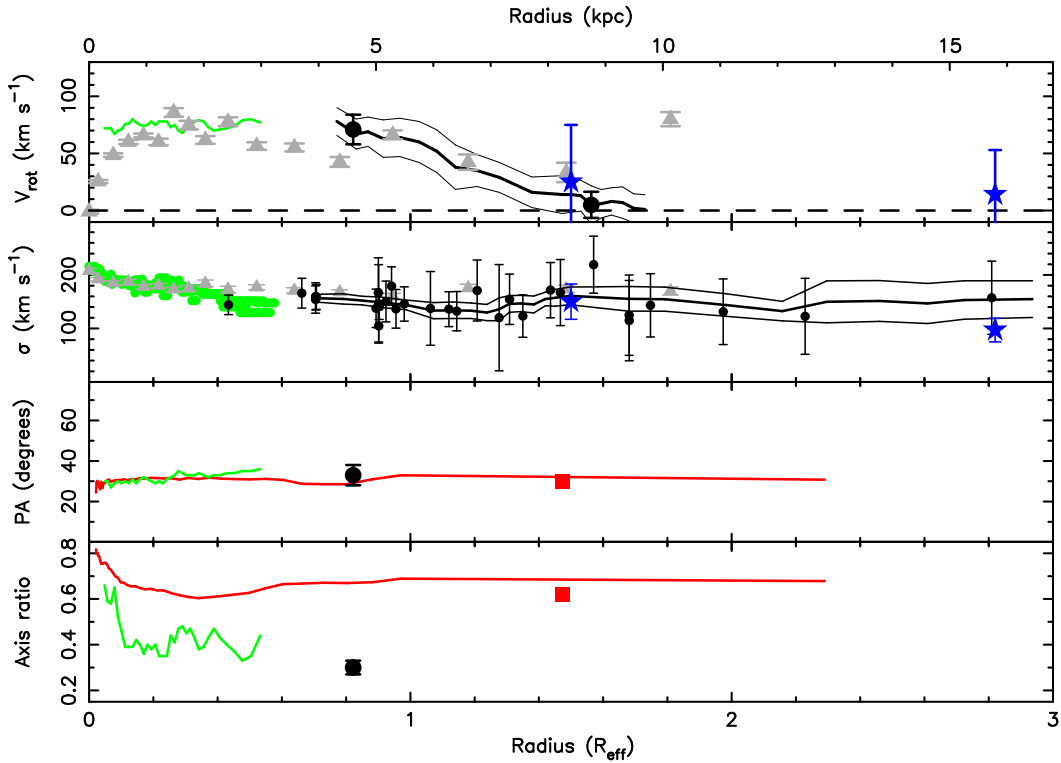}}
\caption{{\bf NGC~821}. Same as Fig. \ref{2768rad} with the following
exceptions. Radial binning is in ranges of R$<$1.0~R$_{\rm eff}$ and
R$\geq$1.0~R$_{\rm eff}$ only. Kinematic PA and axis ratio were only
estimated for the inner range. Inner range values of these parameters
were \emph{assumed} for the measurement of V$_{\rm rot}$. Green data
points are from SAURON (Emsellem et al. 2004). Grey data points are
from Forestell \& Gebhardt (2008). Blue stars are the planetary nebula
data of Coccato et al. (2009). Red lines in PA and axis ratio plots
are the photometric data of Goudfrooij et al. (1994).}
\label{821rad}
\end{figure*}

Rolling fits of the rotation velocity data (Fig. \ref{821rad}) were
performed using 7 data points in the inner region rising to 12 data
points in the outer regions of the galaxy.  Fig. \ref{821rad} confirms
the falling V$_{\rm rot}$ with radius and flat $\sigma$ identified in
the plots with PA. The figure also show that, where they overlap,
there is good agreement between our results and both the (Emsellem et
al. 2004) and Forestell \& Gebhardt (2008) results for V$_{\rm rot}$,
$\sigma$, PA$_{\rm kin}$ and kinematic axis ratio, although we note
one discrepant data point of Forestell \& Gebhardt at 1.8 ~R$_{\rm  eff}$.

Our $\sigma$ values are also in good agreement with literature
studies of both stellar and planetary nebulae kinematics. We note that
Coccato et al. (2009) also measure a planetary nebula rotation
velocity consistent with zero for a data point at 5~R$_{\rm eff}$ with
rotation and $\sigma$ of 14$\pm$25 km s$^{-1}$ and 51$\pm$18 km s$^{-1}$, 
respectively.

The single kinematic axis ratio value that we derive is significantly
lower than the photometric axis ratio, confirming the result found using
the SAURON data (Krajnovi{\'c} 2008).

It is also interesting to note that the extent of the disc evident in
the V$_{\rm rot}$ plot of Fig. \ref{821rad} is similar to the extent
of both the dip in photometric axis ratio of Goudfrooij et al. (1994)
(bottom plot; Fig. \ref{821rad}) \emph{and} to the extent of the young
central population identified in Proctor et al. (2005), both of which
are evident out to nearly 1~R$_{\rm eff}$. This lends credence to the
speculation that the young central ages are associated with the
presence of a disc that has undergone a recent burst of star
formation. Also of interest is the fact that the photometric axis
ratio of the galaxy remains fairly low ($\sim$0.7) even at radii well
beyond the extent of the disc. This suggests that the galaxy is
comprised of a disc embedded in an highly elliptical bulge in
agreement with the photometric classification in the literature
(Michard \& Marchal 1994; Ravindranath et al. 2002; Emsellem et
al. 2004).

\begin{figure*}
\centerline{\psfig{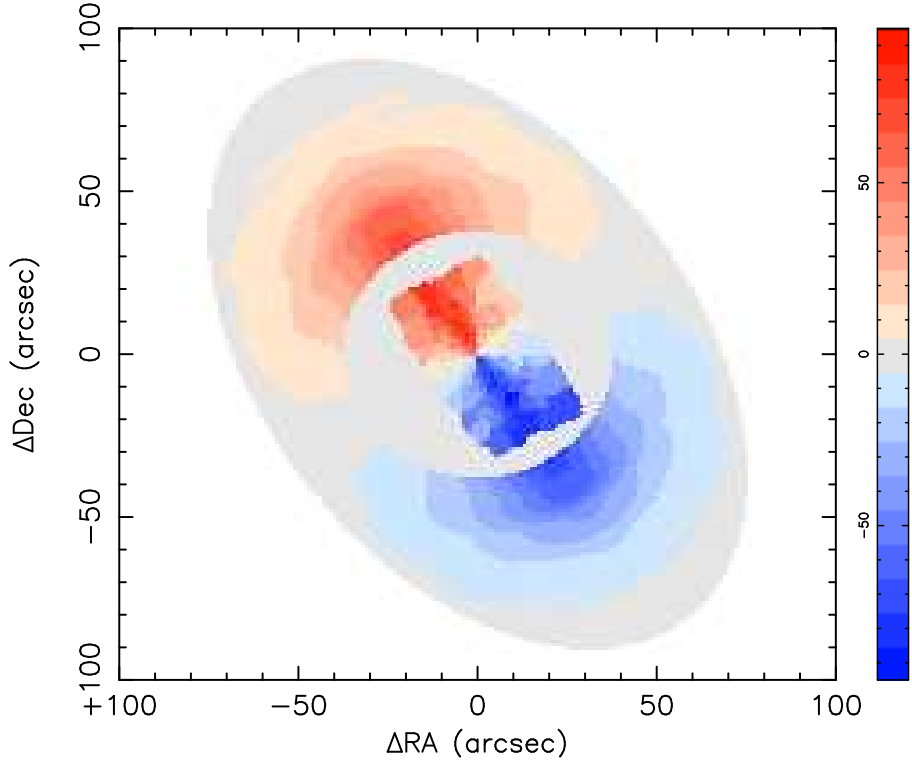}}
\caption{{\bf NGC~821}. Reconstruction of the rotation profile using
  the rolling fits of V$_{\rm rot}$ (Fig. \ref{821rad}). However,
  beyond 1~R$_{\rm eff}$ the values of kinematic PA and axis ratio are
  \emph{assumed} to be the same as within 1~R$_{\rm eff}$ (33\degrees
  and 0.3 respectively).  The outer boundary of the map has been
  chosen to reflect the photometric axis ratio of the galaxy
  (0.62). The colour scale (in km s$^{-1}$) is shown on the right.
  Also shown is the SAURON (Emsellem et al. 2004) map of the central
  regions of the galaxy. The agreement between the data sets is again
  extremely good.}
\label{821image}
\end{figure*}

To aid in the visualisation of our results, and to facilitate the
comparison of our results with those of SAURON (Emsellem et al. 2004),
we have reconstructed the 2-D rotation map of the galaxy in
Fig. \ref{821image}. The image was generated using the rolling fit
values of V$_{\rm rot}$ of Fig. \ref{821rad} and assuming rotation to
have constant PA$_{\rm kin}$ and axis ratio, with values equal to the
values within 1~R$_{\rm eff}$ (33\degrees and 0.3 respectively; Table
\ref{fits}). The SAURON map is also shown.  Despite the spatial gap
between the two data sets, agreement can be seen to be good.

\section{Discussion}
\label{discuss}
We have demonstrated that our analysis gives good agreement with the
SAURON (Emsellem et al. 2004) study and other data from the literature in regions where
they overlap.  We have also shown that the results provide valuable
insights into individual galaxies. We next consider our results in a
more general sense.

\subsection{Rotation}
The five galaxies in our sample exhibit a range of rotation profiles.
Two galaxies (NGC~1407 and NGC~4494) exhibit flat rotation profiles
(within errors) beyond 0.2 and 0.4~R$_{\rm eff}$ respectively. One
galaxy (NGC~2768) exhibits rotation that increases in amplitude over
the full extent of our data (i.e. to $\sim$3~R$_{\rm eff}$). The
remaining two galaxies (NGC~821 and NGC~1400) exhibit rotation curves
that decline beyond 1~R$_{\rm eff}$, and in the case of NGC~821,
becomes consistent with zero rotation by 1.5~R$_{\rm eff}$. It is
therefore evident from our data that studies confined to the central
regions of galaxies can miss vital aspects of their rotation
properties.

\subsection{Velocity dispersion}
The five galaxies in our sample exhibit only shallow slopes in
$\sigma$ with radius beyond 1~R$_{\rm eff}$. Indeed, in most cases,
results in these regions are consistent with zero slope at the
2--$\sigma$ significance level. Such flat $\sigma$ and rotation
profiles (with the possible exception of NGC~821) represent the
classic signature of a dark matter halo in each galaxy. These data
will therefore prove invaluable to future studies that perform full
dynamical modeling of galaxies.

\subsection{Higher order moments}
We note that all 5 galaxies in this study exhibit non-zero h$_3$
values, the sense of which mirror the rotation velocity. This trend is
identified in the NIR spectra of this work in 3 of the 5 galaxies. In
NGC~1400 it is identified in the re-analysis of the Spolaor et
al. (2008a) optical data, while in NGC 821 it is evident in the optical
data of Forestell \& Gebhardt (2008). When h$_3$ follows the inverse
trend of the rotation velocity it is often taken as the signature of a
cold (rotating) stellar population embedded within a hotter (pressure
supported) system. However, Balcells (1991) shows that this signature
can also be induced by the action of a merger event. We therefore
conclude that all five of the observed galaxies are either merger
remnants or harbour embedded discs.

There do not appear to be any clear trends in
the h$_4$ parameter in our current data. we suspect that this is due
to the low signal-to-noise of the majority of our data.

\subsection{Kinematic and photometric misalignment}
For fast-rotators, the SAURON survey revealed that the kinematic and
photometric position angles are usually well aligned, and that their
axis ratios are similar (Emsellem et al. 2007). This is true for two
of our galaxies (NGC~2768 and NGC~1400). However, in both NGC~4494 and NGC821
there appear to be significant offsets between kinematic and photometric axis
ratios, while for NGC~821 an offset in axis ratio is
detected This result for NGC~821 confirms the SAURON result, despite the fact that we have
only one kinematic data point at large radii.

\subsection{Anisotropy diagram}
\label{vose}
\begin{figure}
\centerline{\psfig{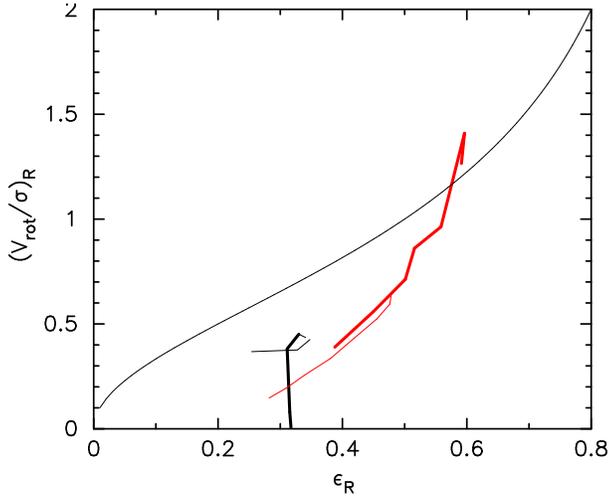}}
\caption{Anisotropy diagram.  V$_{\rm rot}$/$\sigma$ is plotted
  against ellipticity at various radii for NGC~821 (black) and
  NGC~2768 (red). The remaining three galaxies are not shown as their
  V$_{\rm rot}$/$\sigma$ and photometric ellipticities show little or
  no variation with radius. The thin lines represent the inner SAURON
  data, and the thick lines are our data out to 3~R$_{\rm eff}$. The
  smooth curved line represents an ideal isotropic rotator. }
\label{vos_e}
\end{figure}

Binney (1978) showed the usefulness of the anisotropy diagram
(V/$\sigma$ vs photometric ellipticity) as applied to elliptical
galaxies. In Fig. \ref{vos_e} we show this diagram for the two
galaxies in common with the SAURON survey (i.e. NGC 821 and NGC 2768),
with the lines indicating the radial variation.  The plot shows that
the value of V/$\sigma$ in NGC 821 declines with increasing radius,
while NGC 2768 increases. Thus NGC 821 trends away from the isotropic
rotator line, while NGC 2768 trends towards it. It is therefore clear
that the central regions of galaxies do not \emph{necessarily} reflect
the properties of galaxies as a whole, and care must be taken when
drawing conclusions from central data only. We note that similar
conclusions were reached from PN kinematics by Coccato et al. (2009).

\subsection{The SAURON $\lambda_R$ parameter}
As part of the analysis of the SAURON 2-D kinematics, Emsellem et
al. (2007) defined a new parameter, $\lambda_R$ as;

\begin{equation}
\lambda_R = \frac{<R.|V|>}{<R.\sqrt(V^2+\sigma^2)>},
\end{equation}

where $<$R.$|$V$|$$>$ is the luminosity weighted average of the
product of radius and rotation velocity over the surface of the map
and $<R.\sqrt(V^2+\sigma^2$)$>$ is a similarly luminosity weighted
average.  Thus $\lambda_R$ is a measure of the projected angular
momentum per unit mass (i.e. the \emph{specific} angular momentum.

Emsellem et al. (2007) found that galaxies were clearly distinguished
into two classes on the basis of this parameter, i.e.  `fast-rotators'
($\lambda_R$ $>$ 0.1) and `slow-rotators' ($\lambda_R$ $<$ 0.1). Fast
rotators tended to be lower luminosity, discy ellipticals, while the
massive boxy ellipticals were often slow rotators.

The two galaxies in common with the SAURON project (NGC 821 and
NGC~2768) were both classified as `fast-rotators' (i.e. $\lambda_R$
$>$ 0.1) by Emsellem et al. (2007). Their $\lambda_R$ values are very
similar (0.258 and 0.268 respectively). However, we find that when
probed to radii greater than 1 R$_{\rm eff}$ NGC~821 reveals a
declining rotation velocity (and roughly constant velocity dispersion)
profile.  This suggests that if the outer region data were included
the calculated $\lambda_R$ value would be smaller.  Although due to
the luminosity weighting in the $\lambda_R$ definition more emphasis
is given to the central regions, so the actual change in the value
might be small.

\begin{figure}
\centerline{\psfig{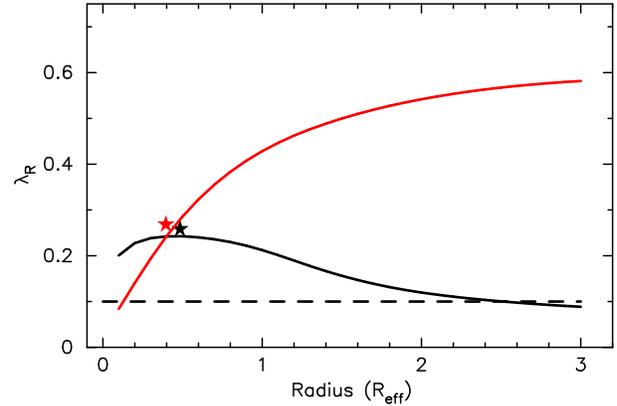}}
\caption{The $\lambda_{\rm R}$ parameter (cumulative value of $\lambda$ within 
radius R) with major axis radius
scaled by the effective radius. The SAURON central values are presented
as stars, and our data by solid lines. Here NGC~821 is black and
NGC~2768 is red.  
The dashed line represents the delimiting value of 0.1 between 'fast-' and
'slow-'rotators (as defined in Emsellem et al. 2007). 
While both galaxies are classified as fast-rotators based
on SAURON data within 1~R$_{\rm eff}$, NGC~821 would be
classified as a slow-rotator if the data 
to 3~R$_{\rm eff}$ are considered.}
\label{lambdar}
\end{figure}

To test this, we constructed a model of each of these galaxies. Each
model assumed R$^{1/4}$ de Vaucouleur's profiles. For each galaxy, the
rotation parameters (V$_{\rm rot}$, PA$_{\rm kin}$ and kinematic axis
ratio) were then characterised by a polynomial fit to the plots with
radius (Figures \ref{2768rad} and \ref{821rad}). The luminosity and
velocity maps so produced were then combined as per Emsellem et al.
(2007) to calculate the $\lambda_R$ values inside ellipses of
increasing effective radius\footnote{Although precise values of the
sersic index for the two galaxies are not available, we note that the
results are relatively insensitive to this parameter}. The results are
presented in Fig. \ref{lambdar}. Agreement with the published SAURON
values for the inner regions can be seen to be extremely
good. However, for NGC~821 our model suggests that, despite its
classification as a fast-rotator (when only data from within 1~R$_{\rm eff}$ 
is considered), the decreasing rotation at large radii could
indeed result in the galaxy being classified as a slow-rotator when
all the data out to 3~R$_{\rm eff}$ are considered.

In contrast to NGC~821, NGC~2768 has a profile that continues to rise
all the way out to $\sim$3~R$_{\rm eff}$. Its $\lambda_R$ value will
therefore continue to increase as larger and larger radii are included
in the calculation, and the galaxy will remain a
`fast-rotator'. Therefore, despite their similarity within one
effective radius, both of these galaxies have vastly different
distributions of angular momentum with radius (see also Coccato et
al. 2009).

Although not observed in the SAURON survey, we assume NGC~1407 to be a
slow-rotator, as its V/$\sigma$ value is low (we estimate 0.09 on the
basis of Equation 23 of Cappellari et al. 2007) and its kinematics are
similar to NGC~5982 which was observed by SAURON (Krajnovi{\'c} et
al. 2008) and defined as a slow-rotator (Emsellem et al.  2007).  In
the case of NGC 1400 and NGC 4494 their relatively high rotation
velocities and relatively low velocity dispersions clearly indicate
them to be fast rotators.

\subsection{The RMS velocity parameter}
In addition to the V$_{\rm rot}$/$\sigma$  
and $\lambda_R$ parameters which probe orbital anisotropy and
angular momentum respectively, another fundamental parameter of a
galaxy is its internal kinematic energy. The observational
measure of the total kinetic energy 
is the RMS velocity parameter, which {\it locally}
is equal to the quadrature combination of rotation and dispersion
velocities, i.e. $V_{\rm rms} = \sqrt{V_{\rm rot}^2 + \sigma^2}$
(e.g. Napolitano et al. 2009). 

\begin{figure}
\centerline{\psfig{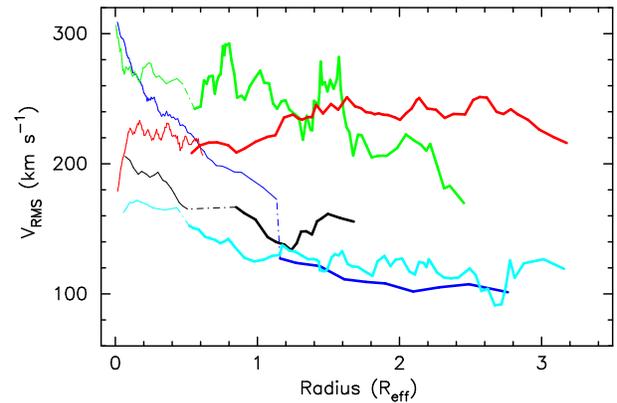}}
\caption{V$_{\rm rms}$ parameter with major axis radius. 
Galaxies are colour coded, i.e 
NGC~821  black, NGC 1400 blue, NGC1407 green, NGC 2768 
red and NGC 4494 cyan. The thin lines represent 
the inner literature data, and the thick lines are our data 
out to larger radii (with extrapolated 
dashed lines connecting the two). In general the 
V$_{\rm rms}$ profiles decline with radius.} 
\label{v2s2}
\end{figure}

In Fig. \ref{v2s2} we show the RMS velocity parameter as a function of
major axis radius for the five galaxies in our sample. We show the
inner regions from literature data and the outer regions from our
data. For all but one galaxy (NGC~2768) a general decline in the RMS
velocity parameter is seen with radius (as might be expected from the
generally flat rotation velocity and declining velocity dispersion
profiles observed). The implications of this for the mass profile of
each galaxy requires detailed dynamical modeling.

\section{Conclusions}
\label{concs}
We have developed a new technique for extracting spectra of the galaxy
background contained in the multi-object observations using 
the DEIMOS spectrograph on the Keck telescope. Here, the slits contained globular
clusters, which were the primary target of the observing programme.
The technique allows measurement of the velocity moments (rotation
velocity, velocity dispersion, and Gauss-Hermite coefficients h$_3$
and h$_4$) in a galaxy halo out to at least 3 effective radii (R$_{\rm
eff}$), covering a surface area of some 24~R$_{\rm eff}^2$, in
comparison to the $\sim$0.4~R$_{\rm eff}^2$ achieved by the SAURON integral
field unit.

For the 5 early-type galaxies in our sample, we have demonstrated that
the agreement between our results and the literature values is
generally extremely good in the regions of overlap, including the
velocity dispersion profiles from planetary nebulae data for NGC 821
and NGC 4494.

For NGC 821 we show that the kinematic signature of the disc
identified in the surface photometry disappears at around 1.5~R$_{\rm
  eff}$, while the photometry remains highly elliptical in the outer
regions. From this we conclude that NGC 821 constitutes a highly
elliptical bulge with an embedded central disc.  For NGC~1400, we also
detect a falling rotation profile beyond 1~R$_{\rm eff}$. When
considered in conjunction with the significant h$_3$ found by Spolaor
et al. (2008a) in the inner regions, this suggests that this early-type
galaxy also contains an embedded disc. For both NGC~1407 and NGC~4494,
which are currently classified as pure elliptical galaxies, the
detection of a trend in the Gauss-Hermite h$_3$ coefficient suggests
that these too harbour a sub-population with `disk-like' kinematics.
In NGC~4494, both the `disk-like' kinematics \emph{and} a coordinated
shift in photometric and kinematic position angles provide strong
support for the idea that this galaxy is a merger remnant. The
kinematics of NGC 2768 clearly indicates that a disc is present.
However, in this case, the disc becomes more prominent with
galacto-centric radius, suggest that this galaxy is, in fact, an S0. 
It is therefore apparent that all five
galaxies in our sample show evidence for a disc or disc-like
components.

We estimate the angular momentum parameter $\lambda_R$ used in the
SAURON survey to classify early-type galaxies as either fast or slow
rotators.  Importantly, our analysis shows that the central kinematics
are not necessarily representative of the kinematics at larger radii,
and that galaxies can change their rotator class when data from larger
radii are considered.  This has important repercussions for any
attempts to classify galaxies by their central kinematic properties
alone.

We have shown that our sample galaxies reveal the relatively flat
rotation and velocity dispersion  profiles suggestive of massive dark
matter halos, thus demonstrating the value that our new technique will
have when full dynamical modeling of these galaxies is carried
out.

Finally, we note that the observations used here were not
optimised for this project. Future work that uses slit masks
designed for the purpose of extracting galaxy halo spectra should
provide additional improvements. \\

\newpage
\noindent{\bf References}\\

\noindent Balcells M., 1991, A\&A, 249, L9\\
Bender R., 1988, A\&A, 193, L7\\
Bender R., 1990, A\&A, 229, 441\\
Bender R., Saglia R.P., Gerhard O.E. 1994, MNRAS 269, 785\\
Binney J., 1978, MNRAS, 183, 501\\
Cappellari M., Emsellem E., 2004, PASP, 116, 138\\
Cappellari M., et al., 2007, MNRAS, 379, 418\\
Carollo, C. M., Franx M., Illingworth G. D., Forbes D. A., 1997, ApJ, 481, 710\\ 
Coccato et al., 2009, MNRAS, in press\\
Da Costa L. N., et al., 1998, AJ, 116, 1\\
Dekel A., Stoehr F., Mamon G. A., Cox T. J., Novak G. S., Primack J. R., 2005, Nature, 437, 707\\
Denicol{\'o} G., Terlevich R., Terlevich E., Forbes D. A., Terlevich A., 2005, MNRAS, 358, 813\\ 
de Vaucouleurs G., 1953, MNRAS, 113, 134\\
de Vaucouleurs  G., de Vaucouleurs  A., Corwin  H. G., Buta  R. J., Paturel  G., Fouque  P., 1991, Third Reference Catalogue of Bright Galaxies ({\bf RC3}). Springer-Verlag,  Berlin, Heidelberg, New York\\
Emsellem  E., et al.,	 2004, MNRAS, 352, 721\\
Emsellem  E., et al., 2007, MNRAS, 379, 401\\
Filippenko A. V., Chornock R., 2000, IAUC, 7511, 2\\
Forbes D. A., Franx Ma., Illingworth G. D., Carollo C. M., 1996, ApJ, 467, 126\\
Forestell A., Gebhardt K., 2008, arXiv:0803.3626  \\
Fried D.L., Illingworth G. D., 1994, AJ, 107, 992\\
Gerhard O. E., 1993, MNRAS, 265, 213\\
Goudfrooij P., Hansen L., J{\o}rgensen H. E., N{\o}rgaard-Nielsen H. U., de Jong T., van den Hoek L. B., 1994, A\&AS, 104, 179\\
Graham A. W., Colless M. M., Busarello G., Zaggia S., Longo G., 1998, A\&AS, 133, 325\\
Hakobyan A. A., Petrosian A. R., McLean B., Kunth D., Allen R. J., Turatto  M., Barbon R, 2008, A\&A, 488, 523\\
Howell J. H., 2005, AJ, 130, 2065\\
Jarrett T. H., Chester T., Cutri R., Schneider S., Skrutskie M., Huchra J. P., 2000, AJ, 119, 2498\\
Jarrett T. H., Chester T., Cutri R., Schneider S. E., Huchra J. P., 2003, AJ, 125, 525\\
Kronawitter A., Saglia R. P., Gerhard O., Bender R., 2000, A\&AS, 144, 53\\ 
Krajnovi{\'c} D., Cappellari M., de Zeeuw P. T., Copin Y., 2006, MNRAS, 366, 787\\
Krajnovi{\'c} D., et al.,  2008, MNRAS, 390, 93\\
Lauer T.R., et al.,  1995, AJ, 110, 2622\\
Longo G., Zaggia S. R., Busarello G., Richter G., 1994, A\&AS, 105, 433\\
Mamon G. A., Lokas E. L. 2005, MNRAS, 363, 705\\
Martel A.R., et al., 2004, AJ, 128, 2758\\
McDermid R. M., et al., 2006, MNRAS, 373, 906\\
McMillan P. J., Athanassoula E., Dehnen W., 2007, MNRAS, 376, 1261\\
Mehlert D., Saglia R. P., Bender R., Wegner G., 2000, A\&AS, 141, 449\\
Michard R., Marchal J., 1994, A\&AS, 105, 481\\
Napolitano N.R., et al., 2009, MNRAS,393, 329\\
Norris M. A., et al., 2008, MNRAS, 385, 40\\
Peletier R. F., Davies R. L., Illingworth G. D., Davis L. E., Cawson M., 1990, AJ, 100, 1091\\
Pinkney J., et al., 2003, ApJ, 596, 903\\
Proctor R. N., Forbes D. A., Forestell A., Gebhardt K.,  2005, MNRAS, 362, 857\\
Proctor R. N., Forbes D. A., Brodie J. P., Strader J., 2008, MNRAS, 385, 1709\\ 
Ravindranath S., Ho L. C., Filippenko A. V.,  2002, ApJ, 566, 801\\
Reda F. M., Proctor R. N., Forbes D. A., Hau G. K. T., Larsen S. S., 2007, MNRAS, 377, 1772\\
Rix H-W., White S. D. M., 1992, MNRAS, 254, 389\\
Romanowsky A. J., et al., 2003, Science, 301, 1696\\
Romanowsky A. J., Strader J., Spitler L., Johnson R., Brodie J. P., Forbes D. A., Ponman T., 2008, ApJ, in press\\
Saglia R.P., 1993, ApJ, 403, 567\\
S{\'a}nchez-Bl{\'a}zquez P., Gorgas J., Cardiel N., 2006, A\&A, 457, 823\\
Sandage A., Tammann G. A., van den Bergh S., 1981, JRASC, 75, 267\\
Sandage  A., Bedke  J., 1994, The Carnegie Atlas of Galaxies. Carnegie Institution of Washington with The Flintridge Foundation,  Washington, DC  \\
Sil'Chenko O.,  2006, ApJ, 641, 229\\
Simkin S. M., 1974, A\&A, 31, 129\\
Spolaor M., Forbes D. A., Hau G. K. T., Proctor R. N., Brough S., 2008a, MNRAS, 385, 667\\
Spolaor M., Forbes D. A., Proctor R. N., Hau G. K. T., Brough S., 2008b, MNRAS, 385, 675\\
Statler T. S., Smecker-Hane T., 1999, AJ, 117,839\\
Thomas J., Jesseit R., Naab T., Saglia R. P., Burkert A., Bender R., 2007, MNRAS, 381, 1672\\
Trentham N., Tully R. B., Mahdavi A.,  2006, MNRAS, 369, 1375\\
van der Marel R. P., 1994, MNRAS, 270, 271\\
van der Marel R. P., Franx, M., 1993, ApJ, 407, 525\\
Weil M. L., Hernquist L., 1996, ApJ, 460, 101\\

\noindent{\bf Acknowledgements}\\ 
The analysis pipeline used in this work to reduce the DEIMOS data was
developed at UC Berkeley with support from NSF grant AST-0071048. We
also use the excellent Penalized Pixel-Fitting method (pPXF) code
developed by Cappellari \& Emsellem (2004) and the data products from
the Two Micron All Sky Survey (2MASS), which is a joint project of the
University of Massachusetts and the Infrared Processing and Analysis
Center/California Institute of Technology, funded by the National
Aeronautics and Space Administration and the National Science
Foundation. The authors acknowledge the data analysis facilities
provided by IRAF, which is distributed by the National Optical
Astronomy Observatories and operated by AURA, Inc., under cooperative
agreement with the National Science Foundation. This work was
supported by the National Science Foundation under Grants AST-0507729
and AST-0808099.  J.S. was supported by NASA through a Hubble
Fellowship, administered by the Space Telescope Science Institute,
which is operated by the Association of Universities for Research in
Astronomy, Incorporated, under NASA contract NAS5-26555. Finally, we
also thank the Australian Research Council for funding that supported
this work.

\newpage

\begin{appendix}
\label{spol}
\section{The re-analysis of NGC~1400 and NGC~1407 long-slit data.}
Spolaor et al. (2008a) used the van der Marel \& Franx (1993) code to measure
velocity moments in 68 and 82 apertures along the major axes for
NGC~1407 and NGC~1400, respectively.  The spectra had a S/N of
$\geq$30~$\AA^{-1}$ at 5000~$\AA$, and reached a radial extent of
0.56~$R_{\rm eff}$ ($\sim$ 4.11~kpc) for NGC~1407 and 1.30~$R_{\rm eff}$
($\sim$ 3.58~kpc) for NGC~1400.  Measurement of moments was then carried
out using 17 template stars of spectral class F5 to K4.  Briefly, the
procedure identified the template that best fit the
central regions (where the signal-to-noise is highest) and applied this
to all spatial slices.  The drawback to this method is that
the best-fitting template was found using only the very central
regions and the existence of strong stellar population gradients
(particularly metallicity) may require differing templates
in the inner and outer regions.

We have therefore re-analysed the data of Spolaor et al. (2008a) using
the pPXF code of Cappellari et al. (2007). The main advantage
of this code is that:

i) A template match is found for \emph{all} individual spatial
slices.

ii) The code finds the \emph{combination} of templates that
best matches the data, further improving the fit.\\

\begin{figure}
\centerline{\psfig{figure=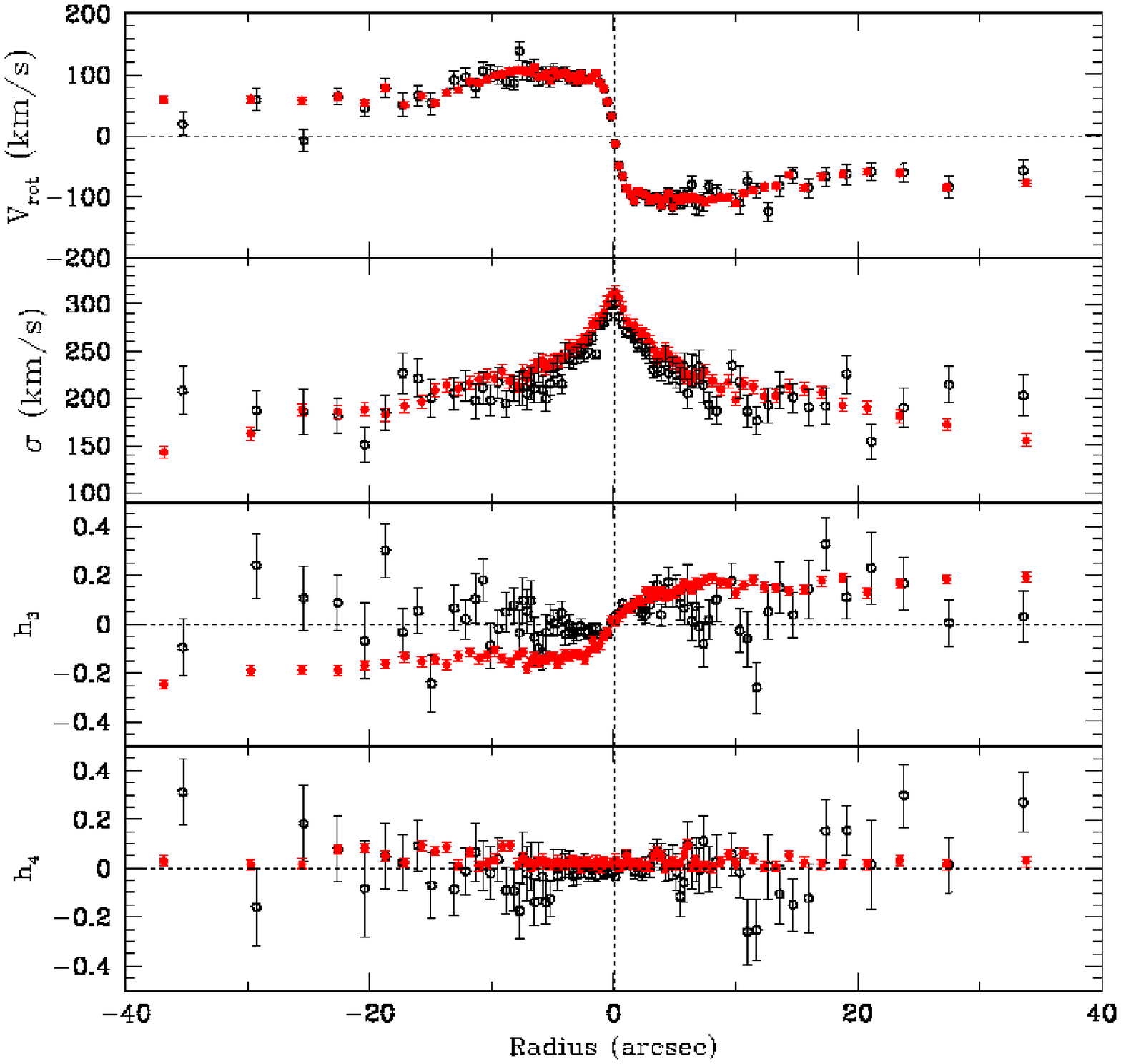,width=9cm}}
\centerline{\psfig{figure=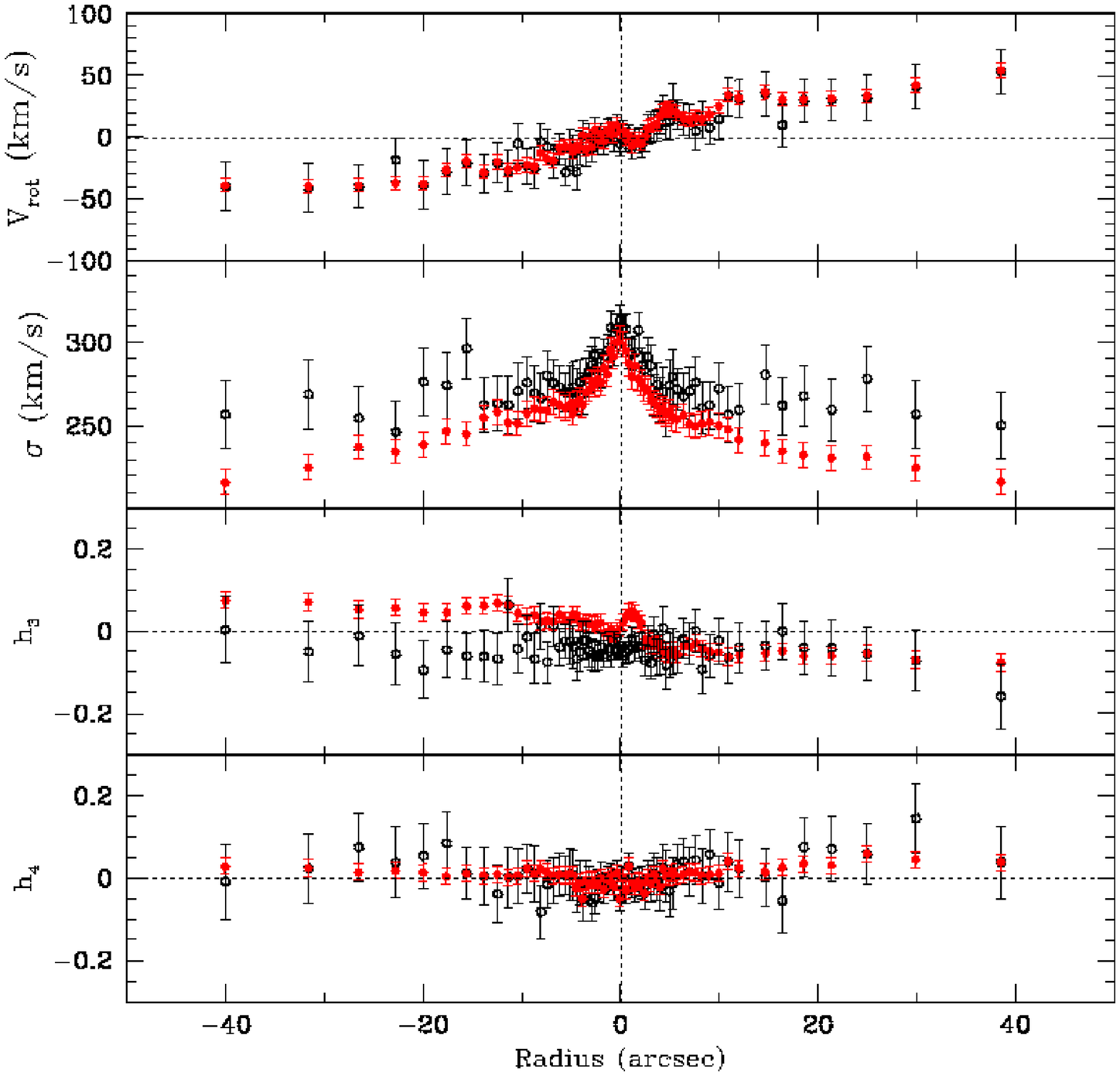,width=9cm}}

\caption{A comparison of the results of the results of Spolaor et
  al. (2008a) (black points; NGC~1400; top and NGC~1407; bottom) to our
  re-analysis using the pPXF software (red points).}

\label{redo}
\end{figure}

The results of the new fits are shown in Fig. \ref{redo} and Table
\ref{apptab}. Errors on derived parameters were estimated by running a
series of Monte Carlo simulations. The results do not change the
general conclusions of Spolaor et al. (2008a). However, the
kinematically decoupled core identified in Spolaor et al. (2008a) is
now much more distinct and the h$_3$ and h$_4$ values appear much more
consistent.

\begin{table}
\begin{tabular}{ccc}
\hline
Parameter & New value & Old value \\
\hline
NGC~1400 & & \\
\hline
V$_{\rm sys}$ & 563 $\pm$ 5  & 560 $\pm$ 10\\
V$_{\rm rot}$  & 113 $\pm$ 6 & 139 $\pm$ 17\\
$\sigma_{0}$& 281 $\pm$ 7  & 301 $\pm$ 4\\
\hline
 & & \\

NGC~1407 & & \\
\hline
 V$_{\rm sys}$ & 1791 $\pm$ 5  & 1794 $\pm$ 10\\
 V$_{\rm rot}$ & 54 $\pm$ 6 & 53 $\pm$ 19\\
$\sigma_{0}$& 270 $\pm$ 7  & 313 $\pm$ 9\\
\hline
\end{tabular} 
\caption{New and old (Spolaor et al. 2008a) kinematic parameters for
NGC~1400 and NGC~1407.}
\label{apptab}
\end{table}

\end{appendix}

\label{lastpage}
\end{document}